\documentclass[floatfix,twocolumn,showpacs,prb,amsmath]{revtex4}
\usepackage{graphicx}
\usepackage{bm}
\usepackage{dcolumn}

\begin{document}
\newcommand{\vn}[1]{{\bf{#1}}}
\newcommand{\vht}[1]{{\boldsymbol{#1}}}
\title{Maximally Localized Wannier Functions within the FLAPW formalism}

\author{F.~Freimuth$^{1}$}
\author{Y.~Mokrousov$^{2}$}
\author{D.~Wortmann$^1$}
\author{S.~Heinze$^2$}
\author{S.~Bl\"ugel$^1$}

\affiliation{$^1$Institut f\"ur Festk\"orperforschung,
Forschungszentrum J\"ulich, D-52425 J\"ulich, Germany}
\affiliation{$^2$Institute for Applied Physics, University
of Hamburg, D-20355 Hamburg, Germany}
\date{\today}
\begin{abstract}
We report on the implementation of the Wannier Functions (WFs) formalism
within the full-potential linearized augmented plane wave method  
(FLAPW),
suitable for bulk, film and one-dimensional geometries.
The details of the implementation, as well as results
for the metallic SrVO$_{3}$, ferroelectric  BaTiO$_{3}$ grown on
SrTiO$_{3}$,
covalently bonded graphene and a one-dimensional Pt-chain are given.
We discuss the effect of spin-orbit coupling on the Wannier Functions
for the cases of SrVO$_{3}$ and platinum. 
The dependency
of the WFs on the choice of the localized trial orbitals
as well as the difference between the maximally localized and
"first-guess" WFs are discussed. Our results on SrVO$_{3}$ and
BaTiO$_{3}$, e.g.\ the ferroelectric polarization of BaTiO$_{3}$, are
compared to results published elsewhere and found to be in excellent
agreement.
\end{abstract}


\maketitle

\section{Introduction}
Commonly, the electronic structure of periodic solids is described
in terms of Bloch functions (BFs), which are eigenfunctions of both the  
Hamiltonian and
lattice translation operators. Due to their delocalized nature BFs
are difficult to visualize and hence do not offer a very intuitive
picture of the underlying physics. Furthermore, BFs do not provide an efficient
framework for the study of local correlations. An alternative  
approach to
electronic structure that does not exhibit these weaknesses is  
provided by
maximally localized Wannier functions (MLWFs). Related to the BFs
via a unitary transformation, MLWFs constitute a mathematically  
equivalent
concept for the study of electronic structure. They are well  
localized in
real space and in contrast to the complex
BFs purely real~\cite{realitynote}. Therefore, it is easy to visualize 
them and to gain
physical insight e.g.\ into the bonding properties of the system under  
study by
extracting characteristic parameters such as the MLWFs' centers, spreads, and
hopping integrals as well as
by analyzing their shapes.

Wannier functions (WFs) were first introduced by Wannier
in 1937~\cite{firstwannier}
as the Fourier transforms of BFs.
Similar to a $\delta$-function, which is the Fourier transform
of a plane wave, WFs are localized in real space while the BFs are
not. However, BFs are only determined up to an arbitrary phase
factor, and hence the definition of WFs as Fourier
transforms of BFs does not specify the WFs uniquely.
As the localization properties of the WFs depend strongly
on the phase factors of the BFs, the Wannier
function approach experienced little enthusiasm until very
recently, after methods for the calculation of WFs with optimal
localization properties had been developed.
One of these new techniques for the construction of localized WFs
is based on the N-th order muffin-tin-orbital 
(NMTO) method.~\cite{WF-NMTO,WannierChemistryPerovskites,WF-NMTO2}
Another method performs at each 
$k$-point a unitary transformation
among the BFs belonging to different bands yielding a new set of  
functions,
the Fourier transforms of which are the MLWFs.~\cite 
{MarzariVanderMLW} The MLWFs
approach is not limited to insulators but is also capable of
providing well localized orbitals for metals.~\cite{entangled}
Only the latter technique is considered in this work.

Sheding new light on otherwise hard to calculate properties of
materials, nowadays MLWFs have almost reached the popularity of
BFs, and using both allows to achieve a
rich diversity in understanding, originating from revealing
both itinerant and localized aspects of electrons in periodic
potentials. For example, a modern theory of
polarization~\cite{polarizationtheory1,polarizationtheory2,cubicBaTiOpol,
RestaPolarization,LayerPolarization}
is based on the displacements of the centers of the MLWFs.
The orbital polarization may be expressed in terms of 
MLWFs.~\cite{OrbPol1,OrbPol2}
Studying the MLWFs for disordered systems yields a
transparent description of bonding properties.~\cite{amorphousSi}
MLWFs provide a minimal basis set
that allows for efficient computations of the quantum transport
of electrons through
nanostructures and molecules.~\cite{transport1,transport2}
Within the research area of strongly correlated electrons
MLWFs are becoming the preferred basis for studying the
local correlations.~\cite{NmtoWanPlusDmft,fullorbitalscheme,dmft}

The MLWFs-induced burst in studying the properties of materials
which are hard to probe on the basis of traditional
band theory is very recent and many subtle aspects, such as
magnetism, various spin-orbit coupling and
non-collinearity-driven effects are still to be put on the
MLWFs footing. In this respect the precision of the computational
electronic structure method used for the construction of the MLWFs
might play a very important role, as sophisticated
details of the electronic structure and tiny energy
scales are involved. In particular in magnetism, the choice of the
appropriate {\em ab initio} method plays a crucial role.
 From this point of view it is common consensus that  the 
full-potential linearized augmented plane wave
method (FLAPW) is one of the most precise electronic structure
methods used today. Ab-initio MLWFs have already been calculated
within the FLAPW framework for MnO~\cite{MnOPosternak} and 
TiO$_{2}$.~\cite{TiOPosternak1,TiOPosternak2}

In the present paper we report in detail on the implementation
of MLWFs within the FLAPW method as implemented in the
{\tt FLEUR}~\cite{fleurcode} code.
The current implementation allows a fast computation of MLWFs 
for a large variety of materials and complex geometries,
including bulk, film~\cite{Krakauer} and truly one-dimensional  
geometrical setups.~\cite{trueonedim} To verify our implementation
we apply the method to four different systems, two different perovskite
systems, SrVO$_3$ and BaTiO$_3$, one metallic and one ferroelectric,
graphene, a covalently bonded material, and a one-dimensional Pt-chain. 
  This article is structured as follows: We start in section II with
a short outline of MLWFs and their construction procedure, defining the
quantities required from the
first-principles calculation based on the density functional theory (DFT)
by the maximal localization algorithm.
First-guess WFs -- originally devised as a starting point for
the MLWF-algorithm, but widely used as a suitable alternative
to the MLWFs -- are introduced.
Then, the details of our FLAPW implementation are described.
In Section III we apply the formalism to SrVO$_3$, BaTiO$_3$,  
graphene and a one-dimensional Pt-chain. We discuss the effects
of spin-orbit coupling on the MLWFs for SrVO$_{3}$ and the Pt-chain.
We compare our results on SrVO$_3$ and BaTiO$_3$ with theoretical
and experimental data, respectively, and find excellent agreement.
Finally we close this work with conclusions in Section IV.

\section{Method}
\subsection{Maximally localized Wannier functions}
For an isolated band, i.e.\ a band that does not become
degenerate with other bands at any $k$-point, with corresponding BFs
$|\psi_{\vn{k}}\rangle$, the definition of WFs
as Fourier transforms of BFs leads to the following
expression:
\begin{equation}\label{singledef}
   |W_{\vn{R}}\rangle=\frac{1}{N}
      \sum_{\vn{k}}e^{-i\vn{k}\cdot\vn{R}}|\psi_{\vn{k}}\rangle,
\end{equation}
where $\vn{R}$ is a direct lattice vector, which specifies
the unit cell the WF belongs to, and the Brillouin
zone is represented by a uniform mesh of $N$ $k$-points.
The $|\psi_{\vn{k}}\rangle$ are normalized with respect to the
unit cell, while the $|W_{\vn{R}}\rangle$ constitute an orthonormal
basis set with respect to the volume of $N$ unit cells.

However, Eq.~(\ref{singledef}) does not define the WFs
uniquely: The BFs are determined only up
to a phase factor -- hence, for a given set of BFs
and a general $k$-point dependent phase $\phi(\vn{k})$,
\begin{equation}
   |W_{\vn{R}}\rangle' = \frac{1}{N}
   \sum_{\vn{k}}e^{-i\vn{k}\cdot\vn{R}}
   e^{i\phi(\vn{k})}|\psi_{\vn{k}}\rangle
\end{equation}
equally constitute a set of WFs. For their use in practice,
it is desirable to have WFs that decay exponentially in real
space, exhibit the symmetry properties of the system studied,
and are real- rather than complex-valued functions~\cite{realitynote}.
For the one-dimensional Schr\"odinger equation and an
isolated single energy band, Kohn~\cite{WannierKohn} has shown
that there exists only one WF which is real~\cite{realitynote}, 
falls off exponentially
with distance and has maximal symmetry. WFs
with maximal spatial localization~\cite{MarzariVanderMLW} (MLWFs)
fulfill these requirements
of real-valuedness~\cite{realitynote}, optimal decay properties 
and maximal symmetry.
The constraint of maximal localization eliminates the nonuniqueness  
of WFs and
determines $\phi(\vn{k})$ up to a constant.

In the general case, energy bands cross or are degenerate at
certain $k$-points, making it necessary to consider
a group of bands. This increases the freedom in
defining WFs further, as now bands may be
mixed at each $k$-point via the
transformation $U_{mn}^{(\vn{k})}$:
\begin{equation}\label{WannierInTermsOfBloch}
   |W_{\vn{R}n}\rangle = \frac{1}{N} \sum_{\vn{k}}
       e^{-i\vn{k}\cdot\vn{R}}
       \sum_{m}U_{mn}^{(\vn{k})}
       |\psi_{\vn{k}m}\rangle,
\end{equation}
where the BF has a band index $m$, the
WF an orbital index $n$, and the
number of bands -- which may depend on the $k$-point --
has to be larger than or equal to the number
of WFs that are supposed to be extracted.
Imposing the constraint
of maximal spatial localization on the WFs determines the
set of $U_{mn}^{(\vn{k})}$-matrices up to a common global
phase.~\cite{MarzariVanderMLW,entangled} In case the number of
bands is equal to the number of WFs, the $U_{mn}^{(\vn{k})}$
matrices are unitary. This situation usually occurs when an
isolated group of bands may efficiently be chosen for the system
under study. In the more general case of entangled energy
bands,~\cite{entangled} however, the number of bands is
$k$-point dependent and $U_{mn}^{(\vn{k})}$ no longer unitary.

\subsection{Maximal localization procedure}
Requiring the spread of the WFs to be minimal imposes
the constraint of maximal spatial localization.
For the spread of the WFs the sum of the
second moments,
\begin{equation}\label{totspreadfunc}
   \Omega = \sum_{n}[\langle \vn{x}^2\rangle_{n}-
                           (\langle\vn{x}\rangle_{n})^2],
\end{equation}
is used,
where $\langle\rangle_{n}$ denotes the expectation value with
respect to the Wannier orbital $|W_{\vn{0}n}\rangle$
and the sum includes all WFs formed from the
composite group of bands. Minimization of the spread
yields the set of optimal $U_{mn}^{(\vn{k})}$-matrices.

An efficient algorithm for the minimization of the spread
Eq.~(\ref{totspreadfunc}) has been given by Marzari and
Vanderbilt first for isolated groups
of bands,~\cite{MarzariVanderMLW} and later on generalized
for the case of entangled energy bands.~\cite{entangled}
The corresponding computer code is publicly available~\cite{MLWcode}
and was used in this work.
Two quantities are required as input by this computational method
and have to be provided by the first-principles calculation:
First, the projections
$A_{mn}^{(\vn{k})}=\langle\psi_{\vn{k}m}|g_{n}\rangle$
of localized orbitals $|g_n\rangle$ onto the BFs
are needed to construct a starting point for the
iterative optimization of the MLWFs.
Second, the overlaps between the lattice periodic parts
$u_{\vn{k} m}(\vn{x}) = e^{-i\vn{k}\cdot\vn{x}}\psi_{\vn{k} m}(\vn{x})$
of the BFs at nearest-neighbor  $k$-points $\vn{k}$ and $\vn{k}+\vn{b}$,
$M_{mn}^{(\vn{k},\vn{b})} =
\langle u_{\vn{k} m}|u_{\vn{k}+\vn{b}, n}\rangle$, 
are necessary to evaluate the relevant 
observables~\cite{MarzariVanderMLW}:
\begin{equation}\label{observablex}
\langle\vn{x}\rangle_{n}=-\frac{1}{N}\sum_{\vn{k},\vn{b}}w_{\vn{b}}\,
\vn{b}\,\Im\ln \tilde{M}_{nn}^{(\vn{k},\vn{b})}
\end{equation}
and
\begin{equation}\label{observablexx}
\langle \vn{x}^{2}\rangle_{n}=\frac{1}{N}\sum_{\vn{k},\vn{b}}w_{\vn 
{b}}\,[1-|\tilde{M}_{nn}^{(\vn{k},\vn{b})}|^2+(\Im\ln \tilde{M}_
{nn}^{(\vn{k},\vn{b})})^2],
\end{equation}
where $w_{\vn{b}}$ is a weight associated with $\vn{b}$,  
and
\begin{equation}\label{mmntilde}
\tilde{M}_{mn}^{(\vn{k},\vn{b})}=\sum_{m_{1}}\sum_{m_{2}}(U_{m_{1}m}^ 
{(\vn{k})})^{*}U_{m_{2}n}^{(\vn{k}+\vn{b})}M_{m_{1}m_{2}}^{(\vn{k},\vn 
{b})}
\end{equation}
evolves during the minimization process due to the iterative  
refinement of the $U_{mn}^{(\vn{k})}$.
The relations Eqns.~(\ref{observablex}, \ref{observablexx})
are valid for uniform $k$-point grids, while in the
continuum-limit the $k$-space expressions for the matrix
elements of the position operator are given by~\cite{MarzariVanderMLW} 
\begin{equation}\label{cenfromwan}
\langle W_{\vn{R}n}|\vn{x}|W_{\vn{0}m}\rangle=
i\frac{V}{2\pi^{3}}\int d^3 k e^{i\vn{k}\cdot \vn{R}}\langle 
\tilde{u}_{\vn{k}n}|\nabla_{\vn{k}}|\tilde{u}_{\vn{k}m}\rangle
\end{equation}
and
\begin{equation}\label{spreadfromwan}
\langle W_{\vn{R}n}|\vn{x}^{2}|W_{\vn{0}m}\rangle=
-\frac{V}{2\pi^{3}}\int d^3 k e^{i\vn{k}\cdot \vn{R}}\langle 
\tilde{u}_{\vn{k}n}|\nabla^{2}_{\vn{k}}|\tilde{u}_{\vn{k}m}\rangle.
\end{equation}
Replacing the gradient $\nabla_{\vn{k}}$ by finite-difference expressions
valid on a uniform $k$-point mesh, one obtains the weights 
$w_{\vn{b}}$ in Eqns.~(\ref{observablex}, \ref{observablexx}).
Through Eqns.~(\ref{observablex}, \ref{observablexx}, \ref{mmntilde})  
the spread $\Omega$ in Eq.~(\ref{totspreadfunc})
may be expressed in terms of and be minimized with respect to the $U_ 
{mn}^{(\vn{k})}$-matrices.

\subsection{First-guess Wannier functions}
The iterative optimization process requires as a starting point
first guesses for the MLWFs.
In order to construct these,
one projects localized orbitals $|g_{n}\rangle$ onto the BF-subspace:
\begin{equation}
   |\phi_{\vn{k}n}\rangle=\sum_{m}|\psi_{\vn{k}m}\rangle\langle
         \psi_{\vn{k}m}|g_{n}\rangle=\sum_{m}A^
         {(\vn{k})}_{mn}\,|\psi_{\vn{k}m}\rangle.
\end{equation}
As the first-guess WFs are supposed to constitute
an orthonormal basis set, the $|\phi_{\vn{k}n}\rangle$ are  
orthonormalized
via the overlap matrix $S_{mn}^{(\vn{k})}=\langle\phi_{\vn{k}m}|\phi_ 
{\vn{k}n}\rangle$
\begin{equation}
|\tilde{\psi}_{\vn{k}n}\rangle=\sum_{m}((S^{(\vn{k})})^{-\frac{1}{2}}) 
_{mn}
|\phi_{\vn{k}m}\rangle,
\end{equation}
before the WFs are calculated from them
\begin{equation}
\label{firstguess}
|W_{\vn{R}n}\rangle=\frac{1}{N}\sum_{\vn{k}}e^{-i\vn{k}\cdot\vn{R}}
|\tilde{\psi}_{\vn{k}n}\rangle.
\end{equation}
While the first-guess WFs are dependent on the choice of localized
orbitals $|g_n\rangle$ they converge to the one and only one
set of MLWFs in the course of the minimization procedure.

Although the first-guess WFs of Eq.~(\ref{firstguess}) are not unique
they agree well with the MLWFs in many cases. Examples where there
is substantial difference between first-guess WFs and
MLWFs include systems where the centers
of the Wannier orbitals do not coincide with the centers of the atoms.
If for the system under study the first-guess WFs are already
satisfactory, one may skip the localization procedure and take
Eq.~(\ref{firstguess}) as the final result.
Computing WFs in such a way
requires much less time, as the $M_{mn}^{(\vn{k},\vn{b})}$ matrix
elements do not have to be calculated and the minimization
of the spread functional is not performed.  First-guess WFs have been
successfully applied to SrVO$_{3}$,~\cite{fullorbitalscheme}
V$_{2}$O$_{3}$~\cite{fullorbitalscheme} and NiO,~\cite{nio} for example.
\subsection{Calculation of $M_{mn}^{(\vn{k},\vn{b})}$
                              within the FLAPW formalism}
For the calculation of MLWFs
the most important quantity
is the $M_{mn}^{(\vn{k},\vn{b})}$ matrix, which
-- according to Eqns.~(\ref{observablex}, \ref{observablexx}) -- contains
all information needed to determine spreads and centers.
With the lattice periodic part $u_{\vn{k} m}(\vn{x})$ being related  
to its
BF
by $u_{\vn{k} m}(\vn{x})=e^{-i\vn{k}\cdot\vn{x}}\psi_{\vn{k} m}(\vn 
{x})$,
the $M_{mn}^{(\vn{k},\vn{b})}$ matrix elements assume the form
\begin{equation}\label{mmnwhatis}
M_{mn}^{(\vn{k},\vn{b})}=\int e^{-i\vn{b}\cdot\vn{x}}(\psi_{\vn{k}m} 
(\vn{x}))^{*}\psi_{[\vn{k}+\vn{b}],n}(\vn{x})\,d^{3}x.
\end{equation}
By $[\vn{k}]$ we denote the wave vector obtained from $\vn{k}$ by  
subtracting the
reciprocal lattice vector that moves $\vn{k}$ into the first Brillouin
zone, according to $[\vn{k}]=\vn{k}-\mathbf{G}(\vn{k})$.

Within FLAPW,~\cite{Hamann:79.1,flapw1} space is partitioned into the
muffin-tin (MT) spheres centered around atoms $\mu$ and the 
interstitial (INT) region. Consequently, 
$M_{mn}^{(\vn{k},\vn{b})}$ has contributions from both,
\begin{equation}
   M_{mn}^{(\vn{k},\vn{b})} = M_{mn}^{(\vn{k},\vn{b})}|_{\text{INT}}+
            \sum_{\mu}M_{mn}^{(\vn{k},\vn{b})}|_{\text{MT}^\mu},
\end{equation}
which will be discussed separately in the following. The treatment of the
vacuum regions occurring in film and one-dimensional setups is discussed in the
appendices \ref{vacappendfilm} and \ref{vacappendonedim}, respectively.

Inside the muffin-tin, the BF is expanded into spherical harmonics,
radial basis functions $u_{l}$, which are solutions of the
scalar relativistic equation at band-averaged energies, and the
energy derivatives $\dot{u}_{l}$
of the $u_{l}$:
\begin{equation}\label{mtexpand}
\begin{aligned}
   &\psi_{\vn{k}m}(\vn{x})|_{\text{MT}^{\mu}} \\
            &=\sum_{L}(A^{\mu}_{L,m}(\vn{k})
   u_{l}^{\mu}(r)+B^{\mu}_{L,m}(\vn{k})
   \dot{u}_{l}^{\mu}(r))
   Y_{L}(\hat{\vn{r}}),
\end{aligned}
\end{equation}
where atom $\mu$ is located at $\vht{\tau}_{\mu}$ and
$\vn{r}=\vn{x}-\vht{\tau}_{\mu}$. Here, $m$ is the band-index and 
$L=(l,l_{z})$ stands for the
angular momentum quantum numbers $l$ and $l_{z}$. The case where
the lapw basis is supplemented with local orbitals is treated
in the appendix \ref{locorbs}.  Using the Rayleigh plane wave expansion
\begin{equation}
e^{-i\vn{b}\cdot\vn{x}}=
4\pi e^{-i\vn{b}\cdot\vht{\tau}_{\mu}}
    \sum_{L}(-1)^{l}i^{l}j_{l}(rb)Y_{L}
    (\hat{\vn{b}})Y_{L}^{*}(\hat{\vn{r}}),
\end{equation}
the contribution $M_{mn}^{(\vn{k},\vn{b})}|_{\text{MT}^{\mu}}$ of the  
muffin-tin
region of atom $\mu$ to the $M_{mn}^{(\vn{k},\vn{b})}$ matrix reads:
\begin{equation}
\begin{aligned}
&M_{mn}^{(\vn{k},\vn{b})}|_{\text{MT}^{\mu}}=4\pi e^{-i\vn{b}\cdot\vht{\tau}_{\mu}}\\
  \times\sum_{L,L'}((&A^{\mu}_{L,m}(\vn{k}))^{*}
         A^{\mu}_{L',n}([\vn{k+b}])t_{11}^{\mu}(\vn{b},L,L')\\
  +(&A^{\mu}_{L,m}(\vn{k}))^{*}B^{\mu}_{L',n}([\vn{k+b}])
                                 t_{12}^{\mu}(\vn{b},L,L')\\
  +(&B^{\mu}_{L,m}(\vn{k}))^{*}A^{\mu}_{L',n}([\vn{k+b}])
                                 t_{21}^{\mu}(\vn{b},L,L')\\
  +(&B^{\mu}_{L,m}(\vn{k}))^{*}B^{\mu}_{L',n}([\vn{k+b}])
                                 t_{22}^{\mu}(\vn{b},L,L')).
\label{mmn_sum}
\end{aligned}
\end{equation}
The matrix elements $t_{11}^{\mu}(\vn{b},L'',L)$ and 
$t_{12}^{\mu}(\vn{b},L'',L)$ are given by the sums over
radial integrals
\begin{equation}
\label{mmnoo}
\begin{aligned}
&t_{11}^{\mu}(\vn{b},L'',L)\\
&=\sum_{L'}
    \mathsf{G}_{ll'l''}^{mm'm''}(\hat{\vn{b}})&\int
    r^{2}j_{l'}(rb)u_{l}^{\mu}(r)u_{l''}^{\mu}(r)d\,r,\\
&t_{12}^{\mu}(\vn{b},L'',L)\\
&=\sum_{L'}
    \mathsf{G}_{ll'l''}^{mm'm''}(\hat{\vn{b}})&\int
    r^{2}j_{l'}(rb)\dot{u}_{l}^{\mu}(r)u_{l''}^{\mu}(r)d\,r,
\end{aligned}
\end{equation}
and analogously for $t_{21}^{\mu}$ and $t_{22}^{\mu}$, where
\begin{equation}
\mathsf{G}_{ll'l''}^{mm'm''}(\hat{\vn{b}})=
    G_{ll'l''}^{mm'm''}i^{l'}(-1)^{l'}Y_{L'}(\hat{\vn{b}}),
\end{equation}
with the Gaunt coefficients
\begin{equation}
G_{ll'l''}^{mm'm''}=\int Y_{lm}(\hat{\vn{r}})Y^{*}_{l'm'}
(\hat{\vn{r}})Y^{*}_{l''m''}(\hat{\vn{r}})\,d\Omega.
\end{equation}

The quantities defined in Eq.~(\ref{mmnoo}) depend on the vectors $ 
\vn{b}$
joining a given $k$-point to its nearest neighbors.
As a uniform $k$-mesh is used the set of $\vn{b}$ vectors
and hence also the integrals defined in Eq.~(\ref{mmnoo}) are  
independent of the $k$-point.
Thus, the quantities Eq.~(\ref{mmnoo}) have to be
calculated only once.

Employing the expansion of the BF in the interstitial region
\begin{equation}
   \psi_{\vn{k} m}(\vn{x}) = \frac{1}{\sqrt{V}}
   \sum_{\vn{G}}c_{\vn{k}m}(\vn{G})e^{i(\vn{k}+\vn{G})\cdot\vn{x}},
\end{equation}
the INT contribution to the $M_{mn}^{(\vn{k},\vn{b})}$ matrix is  
deduced:
\begin{equation}\label{mmnintpre}
\begin{split}
   M_{mn}^{(\vn{k},\vn{b})}|_{\text{INT}} = \frac{1}{V} 
\sum_{\vn{G},\vn{G'}}
     (c_{\vn{k},m}(\vn{G}))^{*} c_{[\vn{k}+\vn{b}],n}(\vn{G'})\\ 
     \times\int_{\text{INT}}e^{i([\vn{k}+\vn{b}]+\vn{G'})\cdot 
\vn{x}}\,
       e^{-i(\vn{k}+\vn{G})\cdot\vn{x}}\,e^{-i\vn{b}\cdot\vn{x}}\,d^ 
{3}x,
\end{split}
\end{equation}
where the integration stretches over the interstitial only.
Introducing the step function $\Theta(\vn{x})$,
that cuts out the muffin tins,
and its Fourier transform $\Theta_{\vn{G}}$,
Eq.~(\ref{mmnintpre}) can be cast into the final form
\begin{equation}
\begin{aligned}
&M_{mn}^{(\vn{k},\vn{b})}|_{\text{INT}}\\
&=\sum_{\vn{G},\vn{G'}}(c_{\vn 
{k},m}(\vn{G}))^{*}
c_{[\vn{k}+\vn{b}],n}(\vn{G'})\Theta_{\vn{G}(\vn{k}+\vn{b})+\vn{G}-\vn 
{G'}},
\end{aligned}
\end{equation}
where $\vn{G}(\vn{k}+\vn{b})$ denotes the
reciprocal space vector that moves $(\vn{k}+\vn{b})$ into the first  
Brillouin zone,
$[\vn{k}+\vn{b}]=\vn{k}+\vn{b}-\vn{G}(\vn{k}+\vn{b})$.
\subsection{Calculation of $A_{mn}^{(\vn{k})}$ within the FLAPW  
formalism}
For the localized orbitals $|g_{n}\rangle$ required to determine
the first-guess WFs, we mostly use functions
that are zero everywhere in space except in the muffin-tin sphere of
that atom, to which the resulting WF is attributed in this sense.
In practice, this works not only for WFs that are atom-centered
but also for bond-centered ones.
Thus, $g_{n}(\vn{x})$ is given by
\begin{equation}\label{locorb}
g_{n}(\vn{x})=\sum_{L}c_{n,L}\tilde{u}_{l}(r)Y_{L}(\hat{\vn{r}}),
\end{equation}
where $\vn{r}=\vn{x}-\vht{\tau}_{\mu}$ is the position relative
to the center of the atom, to which the first-guess WF is attributed,
and the coefficients $c_{n,L}$ control the angular distribution
of $g_{n}(\vn{x})$.
For the radial part $\tilde{u}_{l}(r)$ of the localized orbital
we use the solution $u_{l}^{\mu}(r)$ of the radial scalar  
relativistic equation
for the actual potential obtained from the first-principles calculation
at an energy corresponding to the bands from which the WF is constructed.
It is also possible to use Gaussians,~\cite{MarzariVanderMLW}
or the radial parts of hydrogenic wave functions for $\tilde{u}_{l}(r)$.
Where angular momentum is concerned in Eq.~(\ref{locorb}), contributions
of different angular momenta have to be summed in the general case
to allow the definition of hybrids such
as $sp^3$ orbitals, while there is only an $l=2$ contribution for
WFs corresponding to $d$ orbitals, for 
example.

For a general radial part $\tilde{u}_{l}(r)$ the projection of the
localized orbital $|g_{n}\rangle$ onto the BF is given by
\begin{equation}\label{amncompli}
\begin{array}{cc}
     A_{mn}^{(\vn{k})}=\displaystyle\sum_{L}c_{n,L}[
        (a_{L,m}^{\mu}(\vn{k}))^{*}\int u_{l}^{\mu}(r)
         \tilde{u}_{l}(r)r^{2}dr\\
         \displaystyle+(b_{L,m}^{\mu}(\vn{k}))^{*}\int
        \dot{u}_{l}^{\mu}(r)\tilde{u}_{l}(r) r^{2}dr],
\end{array}
\end{equation}
where the expansion of the BF given in Eq.~(\ref{mtexpand})
was used.
Choosing $\tilde{u}_{l}(r)=u_{l}^{\mu}(r)$ Eq.~(\ref 
{amncompli})
simplifies as follows:
\begin{equation}
A_{mn}^{(\vn{k})}=\langle\psi_{\vn{k}m}|g_{n}\rangle=\sum_{L}c_{n,L} 
(a_{L,m}^{\mu}(\vn{k}))^{*}.
\end{equation}
In order to construct better first guesses for bond-centered
WFs $|g_{n}\rangle$ may also be constructed as a linear combination
of two localized orbitals - one orbital for each atom participating
in the bond. Calculating the WFs for graphene in the next section we  
proceeded
this way.
\subsection{Wannier Representation of the Hamiltonian}
Formulating the Hamiltonian in terms of WFs is a
particularly useful starting point when
effects of correlation~\cite{dmft,nio,fullorbitalscheme}
are studied by DMFT. Furthermore, the hopping integrals
-- along with the MLWFs' spreads, centers and shapes --
provide intuitive insight into the electronic structure.

Written in terms of BFs the Hamiltonian
$\hat{H}$ assumes the diagonal form
\begin{equation}
   \hat{H}=\frac{1}{N}\sum_{\vn{k},n}\epsilon_{n}(\vn{k})
     |\psi_{\vn{k}n}\rangle \langle\psi_{\vn{k}n}|,
\end{equation}
where $\epsilon_{n}(\vn{k})$ stand for the eigenvalues of $\hat{H}$.
If the number of bands is equal to the number of MLWFs extracted
the $U_{mn}^{(\vn{k})}$-matrices in Eq.~(\ref{WannierInTermsOfBloch})
are unitary.
In this case we
arrive at the equivalent form of the Hamiltonian
\begin{equation}\label{RealSpaceHamil}
\hat{H}=\sum_{\vn{R}_{1}m}\sum_{\vn{R}_{2}m'}H_{m,m'}(\vn{R}_{1}-\vn 
{R}_{2})|W_{\vn{R}_{1}m}\rangle\langle W_{\vn{R}_{2}m'}|,
\end{equation}
where
\begin{equation}\label{hoppses}
\begin{array}{cc}
H_{m,m'}(\vn{R}_{1}-\vn{R}_{2})\\[0.2cm]
=\displaystyle\frac{1}{N}\sum_{\vn{k}n}\epsilon_{n}(\vn{k})\langle
W_{\vn{R}_{1}m}|\psi_{\vn{k}n}\rangle\langle \psi_{\vn{k}n}| W_{\vn{R} 
_{2}m'}\rangle
\\[0.2cm]
=\displaystyle\frac{1}{N}\sum_{\vn{k}n}\epsilon_{n}(\vn{k})e^{i\vn{k} 
\cdot(\vn{R}_{1}-\vn{R}_{2})}
\left(U^{(\vn{k})}_{nm}\right)^{*}U_{nm'}^{(\vn{k})}.
\end{array}
\end{equation}
The hopping integrals $H_{m,m'}(\vn{R}_{1}-\vn{R}_{2})$ quantify the  
hopping
of electrons from Wannier orbital $|W_{\vn{R}_{2}m'}\rangle$ into  
Wannier
orbital $|W_{\vn{R}_{1}m}\rangle$.
\subsection{Spin-orbit coupling}
\label{spinorbitcoupling}
In the case of spin-orbit coupling Eq. (\ref{mmnwhatis}) assumes the form
\begin{equation}
M_{mn}^{\vn{k},\vn{b}}=\sum_{\sigma}\int e^{-i\vn{b}\cdot\vn{x}}
                     (  \psi_{\vn{k}m\sigma}   (\vn{x})   )^{*}
                     \psi_{[\vn{k}+\vn{b}],n \sigma}    (\vn{x})
                      d^{3}x,
\end{equation}
where $\psi_{\vn{k}m\sigma} (\vn{x})$ is the BF with lattice vector $\vn{k}$, 
band index $n$, and spin index $\sigma$. The spin index $\sigma$ refers
to the eigenstates of the projection of the spin-operator onto the
spin-quantization axis. Likewise Eq. (\ref{amncompli}) 
has to be changed into
\begin{equation}
\begin{aligned}
     A_{mn}^{(\vn{k})}=&\sum_{L}\sum_{\sigma}c_{nL\sigma}\\
       \times[(&a_{Lm\sigma}^{\mu}(\vn{k}))^{*}\int u_{l,\sigma}^{\mu}(r)
         \tilde{u}_{l,\sigma}(r)r^{2}dr\\
       +(&b_{Lm\sigma}^{\mu}(\vn{k}))^{*}\int
        \dot{u}_{l,\sigma}^{\mu}(r)\tilde{u}_{l,\sigma}(r) r^{2}dr].
\end{aligned}
\end{equation}
In the regime from weak to modest spin-orbit coupling it is reasonable
to choose the localized orbitals $|g_{n}\rangle$ to be eigenstates of
the projection of the spin-operator onto the spin-quantization axis.
This means that for given $n$ $c_{n L \sigma}$ may differ from zero
only for one spin component $\sigma$. 

Eq. (\ref{RealSpaceHamil}) remains valid in the case of spin-orbit coupling, 
but the matrix elements $H_{m,m'}(\vn{R}_{1}-\vn{R}_{2})$ in 
Eq. (\ref{RealSpaceHamil}) 
correspond to hopping between spinor-valued Wannier orbitals then, where the
two spin-components are given by
\begin{equation}
|W_{\vn{R}m\sigma}\rangle=|\sigma\rangle \langle\sigma|W_{\vn{R}m}\rangle,\, \sigma=\uparrow,\downarrow.
\end{equation}
Alternatively, the hopping matrix elements may be decomposed according
to the spin-channels:
\begin{equation}\label{hoppsesspindecomp}
\begin{aligned}
&H_{m m'}^{\sigma \sigma'}(\vn{R}_{1}-\vn{R}_{2})\\
     &=\frac{1}{N}\sum_{\vn{k}n}\epsilon_{n}(\vn{k})
     \langle W_{\vn{R}_{1}m\sigma} | \Psi_{\vn{k}n} \rangle 
     \langle \Psi_{\vn{k}n} | W_{\vn{R}_{2}m'\sigma'} \rangle\\
     &=\frac{1}{N}\sum_{\vn{k}n}\sum_{n' n''}\epsilon_{n}(\vn{k})
      e^{i\vn{k}\cdot (\vn{R}_{1}-\vn{R}_{2})} \\
      &\times (U_{n'' m}^{(\vn{k})})^{*}
       O_{n'' n \sigma}^{(\vn{k})}
       O_{n n' \sigma '}^{(\vn{k})}
       U_{n' m'}^{(\vn{k})},
\end{aligned}
\end{equation}
where the overlap 
$\langle \Psi_{\vn{k} n \sigma} |\Psi_{\vn{k} n' \sigma} \rangle$ 
is denoted $O_{n n' \sigma}^{(\vn{k})}$.
The corresponding real-space representation of the Hamiltonian is given
by
\begin{equation}
\begin{aligned}
\hat{H}=&
\sum_{\vn{R}_{1}m}\sum_{\vn{R}_{2}m '}\sum_{\sigma,\sigma '}\\
    &H_{m,m'}^{\sigma,\sigma'}(\vn{R}_{1}-\vn{R}_{2})
    |W_{\vn{R}_1 m \sigma}\rangle \langle W_{\vn{R}_2 m' \sigma'}|.
\end{aligned}
\end{equation}
Compared with Eq. (\ref{hoppses}) the decomposition 
Eq. (\ref{hoppsesspindecomp}) of the hopping matrix
elements into spin-channels gives further insight into how the spin-channels
are coupled.

The angular characters of the spin-orbit induced corrections can be understood
easily, by applying the $\hat{\vn{L}}\cdot\hat{\vn{S}}$ operator on the MLWFs that one would 
obtain in a calculation without spin-orbit coupling. It is convenient to make use of the identity
\begin{equation}\label{ltimess} 
\hat{\vn{L}}\cdot\hat{\vn{S}}=
\hat{L}_{z}\hat{S}_{z}+\frac{1}{2}[\hat{L}_{+}\hat{S}_{-}+\hat{L}_{-}\hat{S}_{+}].
\end{equation}
As a detailed example we consider the effect of $\hat{\vn{L}}\cdot\hat{\vn{S}}$ on
\mbox{$|d_{xy}\rangle |\uparrow\rangle$}:
\begin{equation}
\begin{aligned}
&\hat{L}_{z}\hat{S}_{z}|d_{xy}\rangle |\uparrow\rangle=-i|d_{x^{2}-y^{2}}\rangle|\uparrow\rangle\\
&\frac{1}{2}\hat{L}_{+}\hat{S}_{-}|d_{xy}\rangle |\uparrow\rangle =\frac{i}{\sqrt{2}}|Y_{2,-1}\rangle |\downarrow\rangle\\
&=-\frac{i}{2}|d_{xz}\rangle |\downarrow\rangle
-\frac{1}{2}|d_{yz}\rangle|\downarrow\rangle
\end{aligned}
\end{equation}
Hence, the resulting idealized MLWF has an up-component the real part of which is $d_{xy}$ and
the imaginary part of which is $-d_{x^{2}-y^{2}}$. The real part of the down-component is
$-\frac{1}{2}d_{yz}$ while the imaginary part of the down-component is given by $-\frac{1}{2}d_{xz}$.
In Table \ref{soc_angular_deps} we list the results for various angular functions for later
reference in the results section. By
\begin{equation} 
d_{3y^2-r^2}=-\frac{1}{2}d_{3z^2-r^2}-\frac{1}{2}\sqrt{3}d_{x^2-y^2}
\end{equation}  
and
\begin{equation}
d_{x^2-z^2}=\frac{1}{2}d_{x^2-y^2}-\frac{1}{2}\sqrt{3}d_{3z^2-r^2}
\end{equation}
we denote the angular functions obtained by rotating $d_{3z^2-r^2}$ and $d_{x^2-y^2}$ around 
the $x$-axis by an angle of $\frac{\pi}{2}$, respectively.
\begin{table}
\caption{\label{soc_angular_deps}Angular part of idealized spin-orbit coupled MLWFs.
Columns 2,3 and 4: Components of the angular function obtained by applying 
$\hat{\vn{L}}\cdot\hat{\vn{S}}$ to the angular function in column 1.}
\begin{ruledtabular}
\begin{tabular}{c|c|c|c}
  $\uparrow$, real part 
& $\uparrow$, imaginary part 
& $\downarrow$, real part 
& $\downarrow$, imaginary part \\
\hline
$d_{xy}$  &$-d_{x^2-y^2}$ &$-\frac{1}{2}d_{yz}$  &$-\frac{1}{2}d_{xz}$ \\
$d_{xz}$  &$\frac{1}{2}d_{yz}$ &$d_{x^2-z^2}$ &$\frac{1}{2}d_{xy}$\\
$d_{3y^2-r^2}$ &$-\frac{1}{2}\sqrt{3}d_{xy}$ &$0.0$ &$-\frac{1}{2}\sqrt{3}d_{yz}$ \\
$p_{z}$   &$0.0$ &$\frac{1}{2}p_{x}$ &$\frac{1}{2}p_{y}$ \\
\end{tabular}
\end{ruledtabular}
\end{table}

For later reference we consider the example of the Wannier orbital 
\mbox{$d_{xy}|\uparrow\rangle_{\text{sqa}}$},
which is an eigenstate of the projection of the spin operator onto
the spin-quantization axis. If the spin-quantization axis does
not coincide with the $z$-direction, a transformation from the
states $|\sigma\rangle_{\text{sqa}}$ to the basis of eigenstates of the
$z$-component of the spin-operator is required before Eq. (\ref{ltimess})
can be applied. For a general spin-quantization axis specified in terms
of angles $\theta$ and $\phi$ the transformation matrix is given by:
\begin{equation}
\left(
\begin{array}{cc}
\cos   \left( \frac{\theta}{2} \right)    e^{-i\frac{\phi}{2}}
&\sin  \left( \frac{\theta}{2} \right)    e^{-i\frac{\phi}{2}}\\
\sin   \left( \frac{\theta}{2} \right)    e^{i\frac{\phi}{2}}
&-\cos  \left( \frac{\theta}{2} \right)    e^{i\frac{\phi}{2}}\\
\end{array}
\right)
\end{equation}
After application of Eq. (\ref{ltimess}) the states are transformed
back to the original basis. We give the result for the spin-quantization
axis pointing in [111]-direction: 
\begin{equation}\label{dxyin111dir}
\begin{aligned}
&\hat{L}_{z}\hat{S}_{z}d_{xy}|\uparrow\rangle_{\text{sqa}}\\
&=-\frac{i}{\sqrt{3}}d_{x^2-y^2}|\uparrow\rangle_{\text{sqa}}-
i\sqrt{  \frac{2}{3} } d_{x^2-y^2}|\downarrow\rangle_{\text{sqa}}\\
&\frac{1}{2}[\hat{L}_{+}\hat{S}_{-}+\hat{L}_{-}\hat{S}_{+}]d_{xy}|\uparrow\rangle_{\text{sqa}}\\
&=\frac{i}{2}\sqrt{\frac{1}{3}}[d_{yz}-d_{xz}]|\uparrow\rangle_{\text{sqa}}+
\frac{\sqrt{2}}{4}[d_{yz}+d_{xz}]|\downarrow\rangle_{\text{sqa}}\\
&+i\frac{\sqrt{6}}{12}[d_{xz}-d_{yz}]|\downarrow\rangle_{\text{sqa}}.
\end{aligned}
\end{equation}
\section{Results}
We have performed first-principles
calculations within the framework of the density functional theory (DFT) 
applying the generalized gradient approximation (GGA) to the DFT.
SrVO$_3$, and BaTiO$_3$ where 
calculated in the bulk mode of the {\tt FLEUR} program, graphene in the 
film mode. For the calculation of the Pt-chain the one dimensional version
of the program was used.

\subsection{SrVO$_{3}$}
\begin{figure}
\includegraphics[width=8cm]{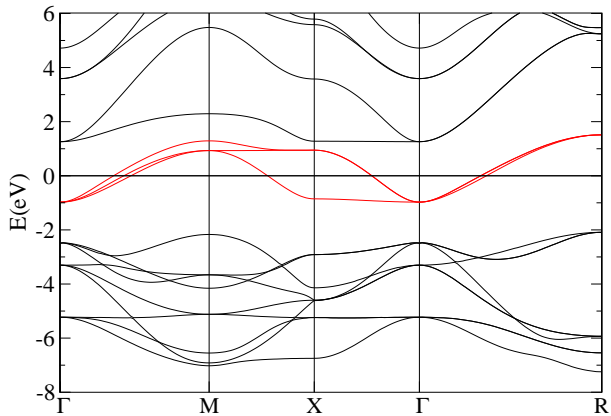}
\caption{\label{srvobands}Bandstructure of SrVO$_{3}$. 
Red: $t_{2g}$-bands around the Fermi level.}
\end{figure}

\begin{figure}
\includegraphics[width=3.5cm]{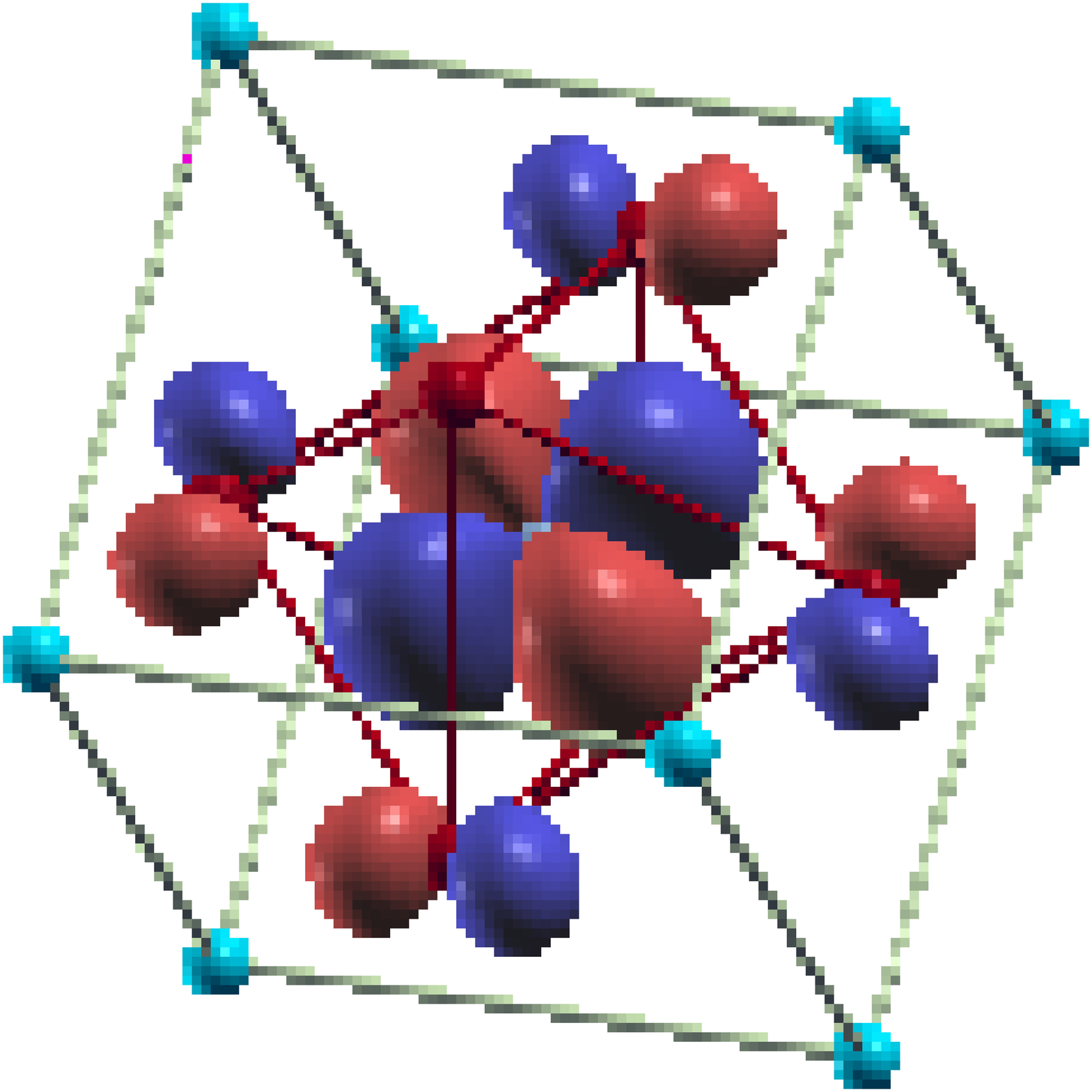}
\includegraphics[width=3.5cm]{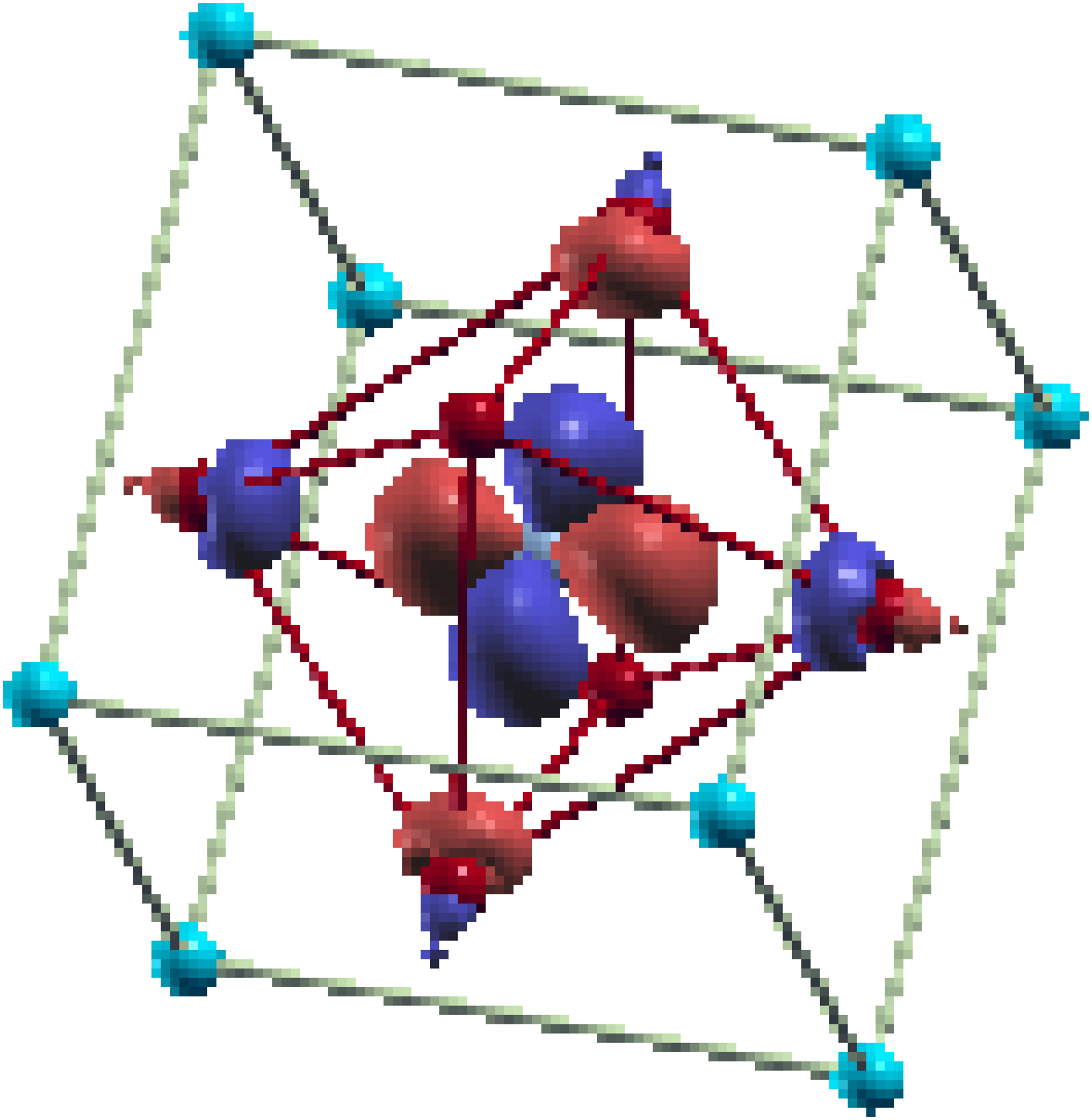}
\caption{\label{srvodxy}Isosurface plot of the $t_{2g}$-like MLWF 
$d_{xy}$ for SrVO$_{3}$ calculated with spin-orbit coupling.
Left: Spin-up component (real part), isosurface=$\pm$0.05.
Right: Spin-down component (imaginary part), isosurface=$\pm$0.001.  
The color of the isosurface refers to
the sign: Positive for dark red and negative for dark blue.
Red balls:
O sites, cyan balls: Sr sites, V site at the center. The WFs were 
plotted using the program XCrySDen~\cite{xcrysden2}.}
\end{figure}

The transition-metal oxide SrVO$_{3}$ crystallizes in a perfectly
cubic perovskite lattice with a lattice constant of 7.26~a.u..
The Sr ions are placed at the corners of a cube (see Fig.~\ref{srvodxy}).
The O ions are placed at the face centers and form an ideal 
octahedron in the center of which the V ion is located.
SrVO$_{3}$ is a metal 
with an isolated group of three $t_{2g}$ bands
around the Fermi level, which are partially occupied by one
$d$-electron (See Figure \ref{srvobands}). Within our GGA calculation
we obtained a bandwidth
of 2.5 eV for the $t_{2g}$-group.
The experimental lattice constant was assumed. We used the
exchange-correlation potential of 
Perdew, Burke and Ernzerhof.~\cite{PerdewBurkeErnzerhof}
For Sr, V, and O muffin-tin radii of 2.8~a.u., 2.1~a.u.\ and
1.4~a.u.\ were used, respectively. Calculations were carried out with
a plane wave cut-off of 4.5~a.u.$^{-1}$.
A uniform 16$\times$16$\times$16
$k$-point mesh was used for the Wannier construction.
For the three $t_{2g}$ bands we constructed three MLWFs,
$d_{xy}$, $d_{yz}$ and $d_{xz}$, which
are equivalent due to symmetry.
The MLWFs are centered at the V site. The spread, Eq.~(\ref 
{totspreadfunc}),
of the MLWFs was found to be 6.97 a.u.$^{2}$ for each of the three  
orbitals.
The first-guess WFs are characterized by a spread which is only 3$\cdot$10$^ 
{-4}$~a.u.$^2$ larger,
showing that MLWFs and first-guess WFs are nearly identical in this  
case.
To investigate the influence of spin-orbit coupling on the 
MLWFs a calculation including spin-orbit coupling was performed for 
the plots (see section \ref{spinorbitcoupling}). The spin-quantization
axis, which defines the two spin-components of the spinor-valued MLWF, 
was chosen in [111] direction, to ensure that the spin components of the
6 spin-orbit MLWFs are related by symmetry. The spin-orbit MLWFs are 
complex-valued. The imaginary parts of the up and down-components of the 
\mbox{$d_{xy}|\uparrow\rangle$}-dominated
orbital, for example, are $d_{x^2-y^2}$-like plus an admixture of
\mbox{$d_{yz}$-$d_{xz}$}, while the real part of the
down-component is \mbox{$(d_{yz}+d_{xz})$}-like. This result can be understood
from the simple model in section \ref{spinorbitcoupling} that leads to 
Eq. (\ref{dxyin111dir}).
The isosurface-plot for the
$d_{xy}$-dominated orbital given in Fig.~\ref{srvodxy} clearly shows the
hybridization between the V($t_{2g}$) and O($2p$) orbitals. 
The symmetry-inequivalent hopping integrals
$H_{m,m'}(\vn{R}_{1}-\vn{R}_{2})$, Eq.~(\ref{hoppses}), are
listed in Table~\ref{srvohoppings} and found to agree well with
recently published WF-results~\cite{dmft,WannierChemistryPerovskites} 
on SrVO$_{3}$. For reasons of
symmetry the 1st-nearest-neighbor hopping integrals between different
orbitals (e.g.\ $d_{xz}$ and $d_{yz}$) are zero in Table~\ref{srvohoppings}. 
However, there is a
coupling between the $d_{xz}$ orbital and the $d_{yz}$ orbitals at the
110 and 111 sites, for example. Due to the dominance of the
nearest-neighbor hopping the three MLWFs may, nevertheless,
approximately be considered independent. The fast decay of the
hoppings with distance furthermore indicates the short-range bonding
in SrVO$_{3}$. The dominance of the 001-hopping for the $d_{xz}$-orbital
over the 010-hopping reflects the restriction of electron hopping to the
$xz$-plane.

In order to study the convergence of the MLWFs with number of $k$-points we
performed a second calculation using an 8$\times$8$\times$8-mesh of $k 
$-points.
This yielded hoppings identical to those of the previous calculation,  
but a
slightly smaller spread of 6.73~a.u.$^{2}$ per orbital. This latter  
difference
is attributed to the fact that the spread was calculated via the  
finite difference
formulae Eqns.~(\ref{observablex}, \ref{observablexx}).

\begin{table}
\caption{\label{srvohoppings}Hopping Integrals for SrVO$_{3}$.  
Energies are in meV.}
\begin{ruledtabular}
\begin{tabular}{l|rrrrrrrr}
$xyz$  &001 &010 &011 &101 &110 &111 &002 &020\\
\hline\\
$d_{xz},d_{xz}$ &$-$262.0 &$-$27.0 &5.8 &$-$84.0 &5.8 &$-$5.7 &7.6 &0.2\\
$d_{xz},d_{yz}$ &0.0    &0.0   &0.0 &0.0   &9.2 &3.6  &0.0 &0.0\\
\end{tabular}
\end{ruledtabular}
\end{table}

\subsection{BaTiO$_{3}$}
As a simple application of the Wannier-function scheme
we present the calculation of the ferroelectric polarization
of the ferroelectric perovskite BaTiO$_3$. 
The evaluation of
the polarization from a DFT calculation of an infinite crystal
can be achieved by means of the Berry-phase technique. After the  
construction
of MLWFs for the occupied valence bands this leads to
the following expression for the polarization~\cite{polarizationtheory1,
polarizationtheory2,cubicBaTiOpol,
RestaPolarization,LayerPolarization}
\begin{equation}\label{electricdipoleformula}
\vn{P}=\sum_{i}q_i\vn{X}_i +\sum_{n}e\langle\vn{x}\rangle_n,
\end{equation}
where $q_i$ and $\vn{X}_i$ denote charge and position of the ion
cores and $\langle\vn{x}\rangle_n$ are the centers of the 
occupied Wannier orbitals.

\begin{figure}
\includegraphics[width=3.5cm]{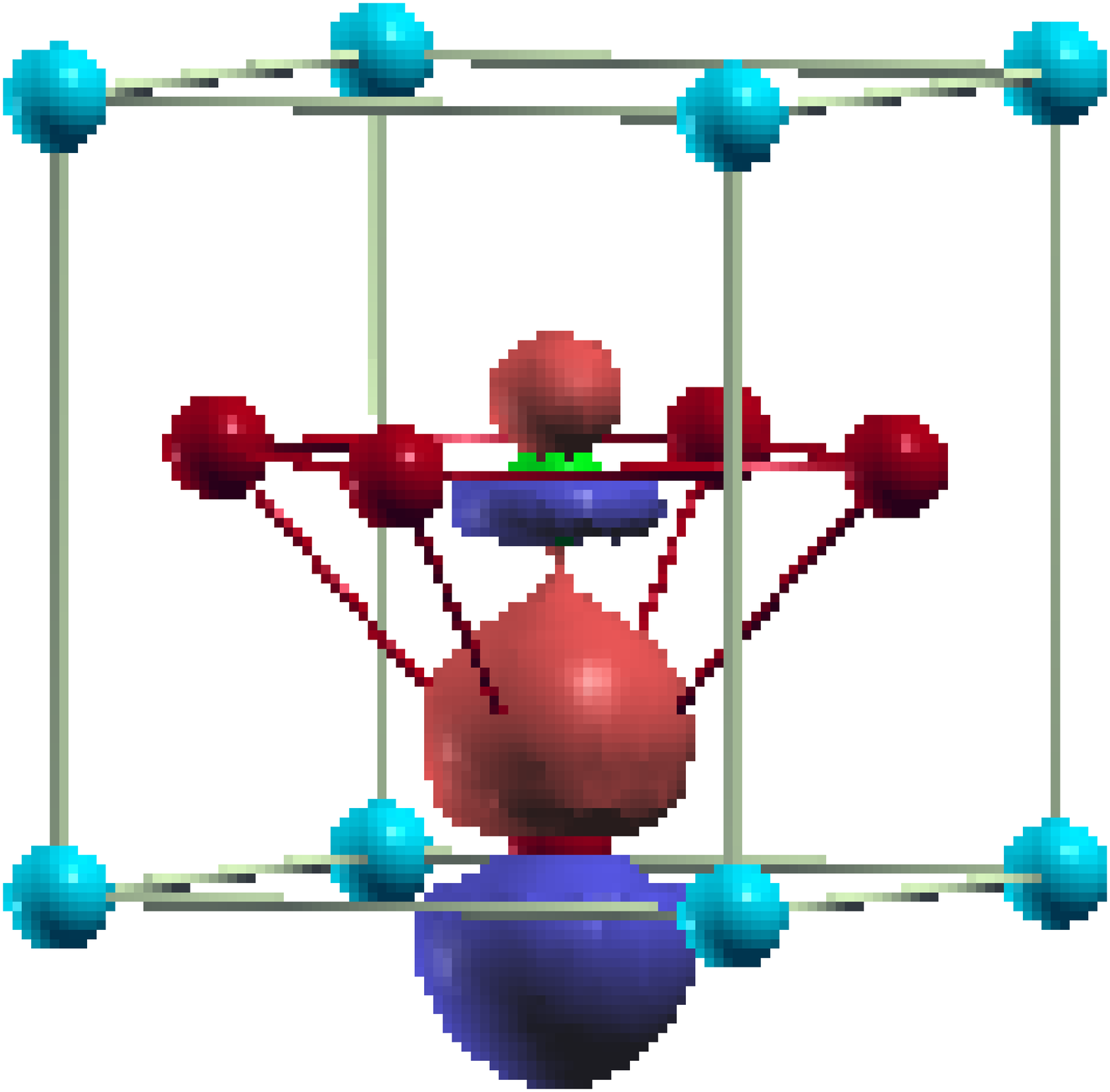}
\includegraphics[width=3.5cm]{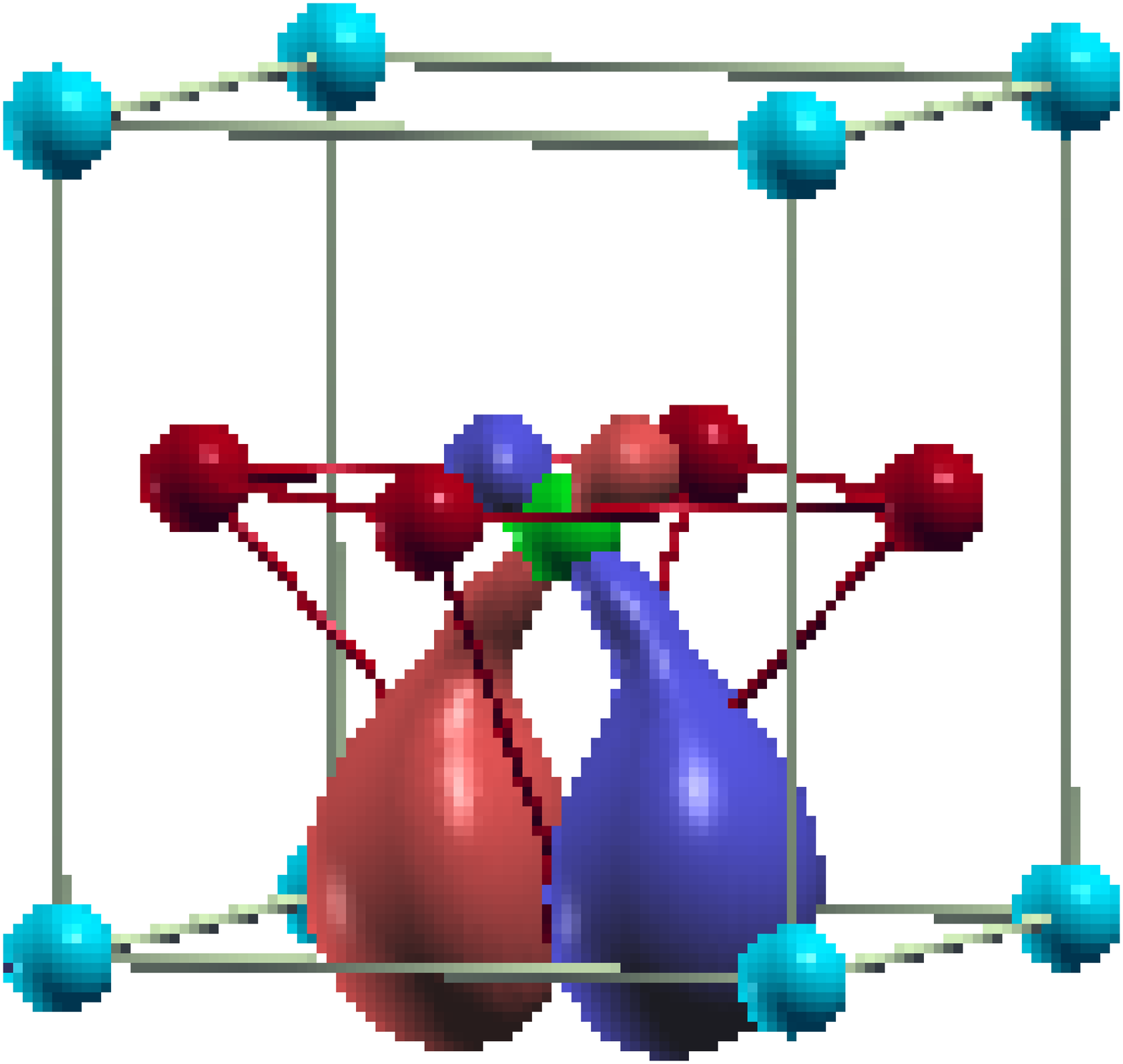}
\caption{\label{batioxy} MLWFs $O(p_{z})$ and $O(p_{y})$ for the oxygen
site close to $xy$-plane in BaTiO$_{3}$. Isosurface=$\pm$0.05. Red balls in the
face centers:
O sites, cyan balls at the corners: Ba sites, green ball at the center:
Ti site. The O site above the upper face of the cuboid is not depicted. }
\end{figure}

\begin{figure}
\includegraphics[width=2.8cm]{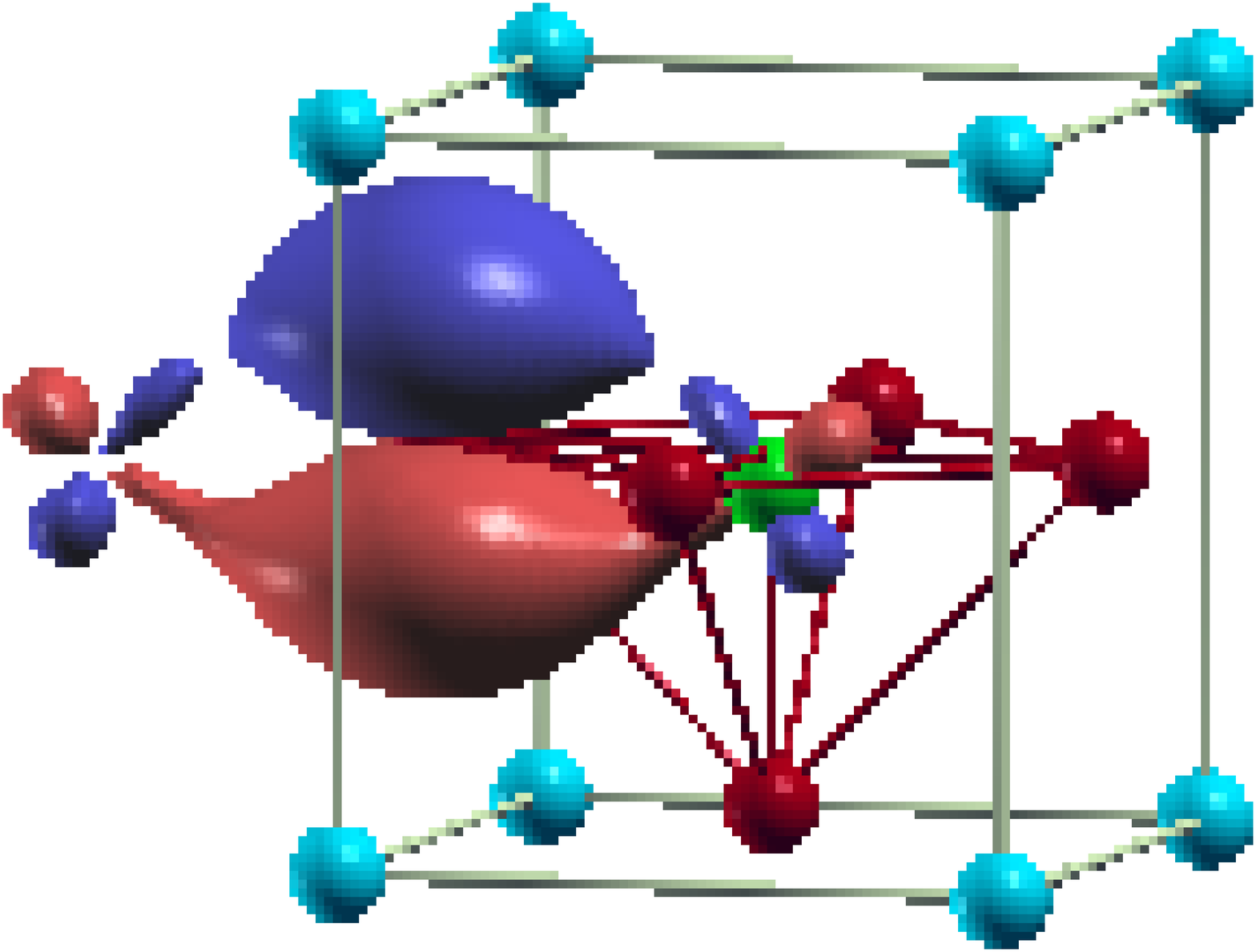}
\includegraphics[width=2.8cm]{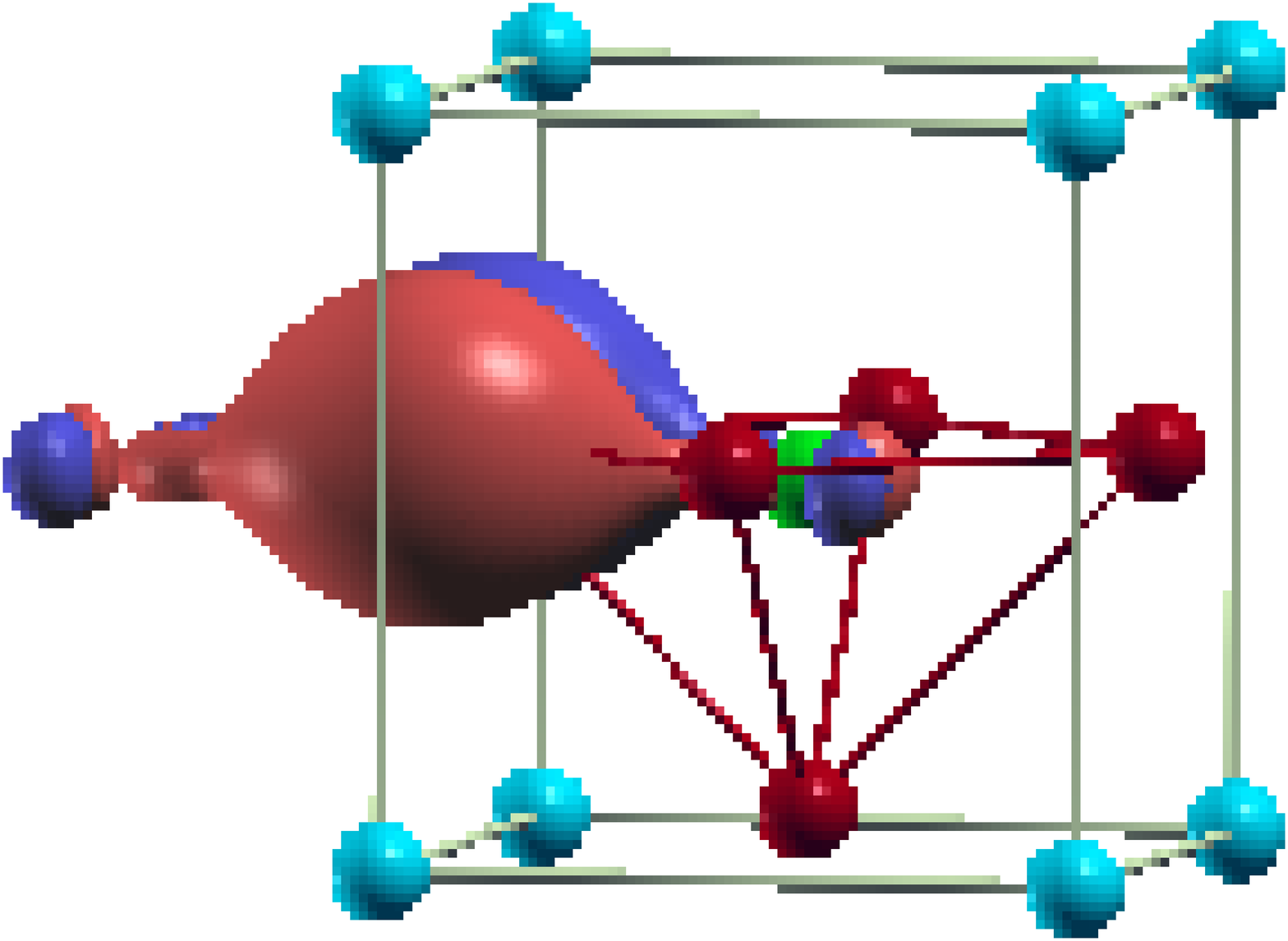}
\includegraphics[width=2.8cm]{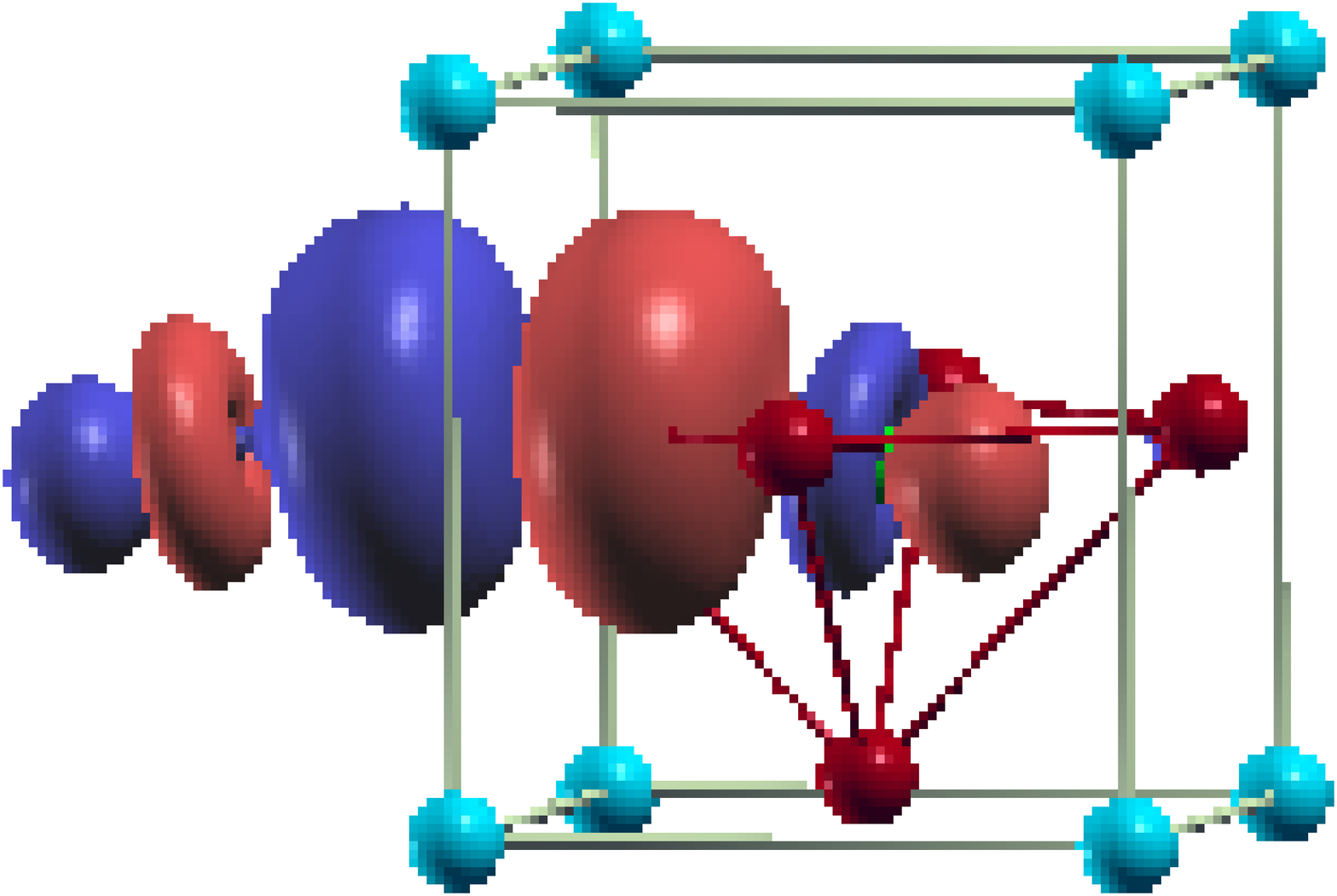}
\caption{\label{batioxz}MLWFs $O(p_{z})$, $O(p_{x})$, and $O(p_{y})$ for
the oxygen site close to $xz$-plane in BaTiO$_{3}$. Isosurface=$\pm$0.05.
Red balls in the
face centers:
O sites, cyan balls at the corners: Ba sites, green ball at the center:
Ti site. }
\end{figure}

We applied this formalism to strained BaTiO$_3$ which is assumed to  
have been
grown epitaxially on top of SrTiO$_3$ assuming the in-plane lattice constant
($a=7.46$~a.u.) of SrTiO$_3$.
We did not consider any finite thickness or interface
effects but simply assumed that this epitaxial relation will hold for 
reasonably thin films.
The lattice constant perpendicular ($c$) as well as the  
positions of all atoms
in the unit-cell where then relaxed by a series of force and total  
energy calculations.
For Ba, Ti and O, muffin-tin radii of 2.2~a.u., 2.0~a.u.\ and 1.3~a.u.\ 
were used, respectively. The plane wave cut-off was chosen to be 4.8 a.u.$^{-1}$.
Using the exchange correlation potential of Perdew and Wang~\cite 
{ExCorPerdewWang}
we obtained a $c/a$ ratio of 1.07, in reasonable agreement with  
experimental data.~\cite{BaTiOEXP}
The resulting atomic positions are given in Table~\ref 
{batiocoordinates} and the crystal structure of  
BaTiO$_{3}$ is illustrated in Figures \ref{batioxy} and \ref{batioxz}.  
Compared to the cubic perovskite structure, the oxygen atoms are moved
out of the face centers and the cube is elongated in $z$-direction.
$z$-reflection symmetry is lost. $\Delta z$ in 
Table~\ref{batiocoordinates} specifies the displacement of the oxygen
and titanium atoms from the symmetric positions in the face centers
and the center of the cuboid, respectively.

We calculated MLWFs separately for the 9 oxygen $p$-bands, the 3  
barium $p$-bands,
the 3 oxygen $s$-bands, the one barium $s$-band, and the 3 titanium $p 
$-bands (the
remaining electrons were treated as core electrons) using a uniform $k 
$-point
mesh of 16$\times$16$\times$16 $k$-points. As final spread, Eq.~(\ref 
{totspreadfunc}),
48.03~a.u.$^{2}$ were obtained for the 9 oxygen $p$ MLWFs while the  
spread of the first-guess WFs was 48.08~a.u.$^{2}$, demonstrating 
that first-guess WFs  and MLWFs are nearly identical for BaTiO$_{3}$.
Figures \ref{batioxy} and
\ref{batioxz} show the isosurfaces of the resulting MLWFs. The MLWFs  
clearly
reflect the broken $z$-reflection symmetry.
Table~\ref{batiowanniercoordinates} lists the coordinates of the  
centers of the
MLWFs along with their deviations $\Delta z$ from the ion sites. As  
evident from there,
the oxygen-MLWFs for the site close to the $xy$-plane exhibit the  
largest
response to the broken $z$-reflection symmetry.
Applying Eq.~(\ref{electricdipoleformula}) we find a polarization of  
48.9~$\mu$C/cm$^2$
in excellent agreement with
experimental data~\cite{BaTiOEXP} of 43~$\mu$C/cm$^2$ for the case of 
thin BaTiO$_3$ layers grown on
SrTiO$_3$. The displacements of the centers of the MLWFs  
with respect to the centers of the atoms contribute 36$\%$ to the polarization.

In order to assess convergence of the results with respect to  the
number of $k$-points a comparative calculation was performed using an  
8$\times$8$\times$8
$k$-point mesh. This calculation yielded a final spread of 
47.19~a.u.$^{2}$ for the MLWFs
of the 9 oxygen $p$ bands and a total polarization
of 48.6~$\mu$C/cm$^2$. We assume these small differences
to be finite difference errors introduced by using
formulae Eqns.~(\ref{observablex}, \ref{observablexx}). 

\begin{table}
\caption{\label{batiocoordinates}Positions of the Ba, Ti and O ions  
in the
constrained ferroelectric perovskite BaTiO$_{3}$ (atomic units).
For the O ions, $\Delta z$ is the displacement from the face centers.
For the Ti ion, $\Delta z$ specifies the displacement from the center
of the cuboid. 
}
\begin{ruledtabular}
\begin{tabular}{l|rrrr}
   &$x$     &$y$    &$z$   &$\Delta z$ \\
\hline\\
Ba &0.000   &0.000  &0.000 &0.000\\
Ti &3.730   &3.730  &3.901 &$-$0.092\\
O  &3.730   &3.730  &0.449 &0.449\\
O  &3.730   &0.000  &4.284 &0.292\\
O  &0.000   &3.730  &4.284 &0.292\\
\end{tabular}
\end{ruledtabular}
\end{table}

\begin{table}
\caption{\label{batiowanniercoordinates}BaTiO$_{3}$: Coordinates,  
displacements 
and spreads
of the MLWFs (atomic units).}
\begin{ruledtabular}
\begin{tabular}{l|rrrrr}
    &$x$          &$y$         &$z$      &$\Delta z$ &$\langle\vn{x}^2 
\rangle$\\
\hline\\
O ($pz$)  &3.730  &3.730  & 0.629 & 0.181    & 4.75\\
O ($px$)  &3.730  &3.730  & 0.686 & 0.238    & 5.69\\
O ($py$)  &3.730  &3.730  & 0.686 & 0.238    & 5.69\\
O ($pz$)  &3.730  &0.000  & 4.296 & 0.012    & 5.69\\
O ($px$)  &3.730  &0.000  & 4.300 & 0.016    & 5.53\\
O ($py$)  &3.730  &0.000  & 4.255 &$-$0.029  & 4.73\\
O ($pz$)  &0.000  &3.730  & 4.296 & 0.012    & 5.69\\
O ($px$)  &0.000  &3.730  & 4.255 &$-$0.029  & 4.73\\
O ($py$)  &0.000  &3.730  & 4.300 & 0.016    & 5.53\\

Ba ($pz$) &0.000  &0.000  &$-$0.047 &$-$0.047  & 6.03\\
Ba ($px$) &0.000  &0.000  &$-$0.011 &$-$0.011  & 6.15\\
Ba ($py$) &0.000  &0.000  &$-$0.011 &$-$0.011  & 6.15\\

O ($s$)   &3.730  &3.730  & 0.542 & 0.095  & 2.77\\
O ($s$)   &3.730  &0.000  & 4.305 & 0.021  & 2.64\\
O ($s$)   &0.000  &3.730  & 4.305 & 0.021  & 2.64\\

Ba ($s$)  &0.000  &0.000  & 0.000 & 0.000  & 3.20\\

Ti ($pz$) &3.730  &3.730  &3.863  &$-$0.038  & 1.48\\
Ti ($px$) &3.730  &3.730  &3.905  & 0.003    & 1.47\\
Ti ($py$) &3.730  &3.730  &3.905  & 0.003    & 1.47\\

\end{tabular}
\end{ruledtabular}
\end{table}


\subsection{Graphene}
Graphene is a covalently bonded system. Consequently,
one expects that the MLWFs are bond centered. This is a particularly 
stringent test for our implementation as the   
LAPW basis functions in which the BFs are expanded (see Eq.~\ref{mtexpand}) 
are centered around the atoms.  
Actually, the four valence bands do not  
constitute
an isolated group of bands as they touch an unoccupied band 
at the $\overline{K}$-point.
Avoiding the $\overline{K}$-point 
when choosing the uniform $k_{\scriptscriptstyle \Vert}$-mesh,  
disentangling is not necessary, however. A single layer of graphene was 
calculated within the {\tt FLEUR} film mode. The muffin-tin radii and
the plane wave cut-off were chosen to be 1.28 a.u.\ and 4.6 a.u.$^{-1}$, respectively.
The C-C bond length was assumed to be 2.72 a.u..
We used the exchage-correlation potential of Perdew, Burke, 
and Ernzerhof.~\cite{PerdewBurkeErnzerhof}
MLWFs and first-guess WFs were constructed for the four
valence bands using an 8$\times$8 $k_{\scriptscriptstyle \Vert}$-mesh
in the two-dimensional Brillouin zone.
For the construction of the first-guess WFs, two calculations were  
performed:
In one calculation the localized functions $|g_{n}\rangle$  
corresponding to
the $sp^{2}$-bonds were chosen to be restricted to the muffin-tin sphere of  
only one atom (FWF1),
while they were restricted in the second calculation (FWF2) to the muffin-tins 
of the two atoms  participating
in the covalent bonding. The FWF2s were nearly
identical with the MLWFs, having the same centers and negligibly  
different spreads,
in particular.
The FWF1s are not centered in the middle of the C-C-bond, the 
FWF2s are, however, centered. Irrespective of the starting point (i.e.\ either
FWF1 or FWF2) 
we arrive at the same MLWFs, which are bond centered.

\begin{figure}
\includegraphics[width=3.5cm]{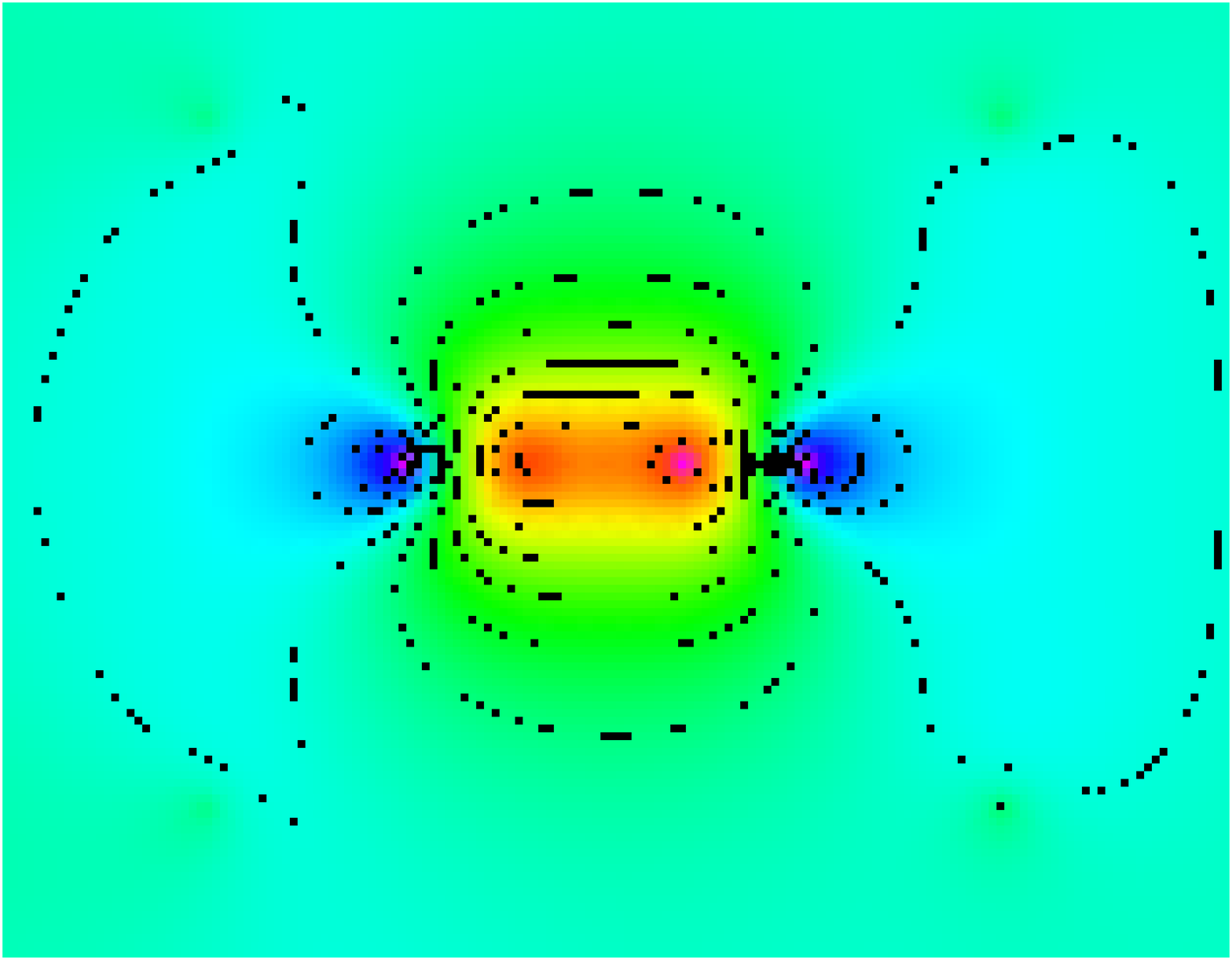}
\includegraphics[width=3.5cm]{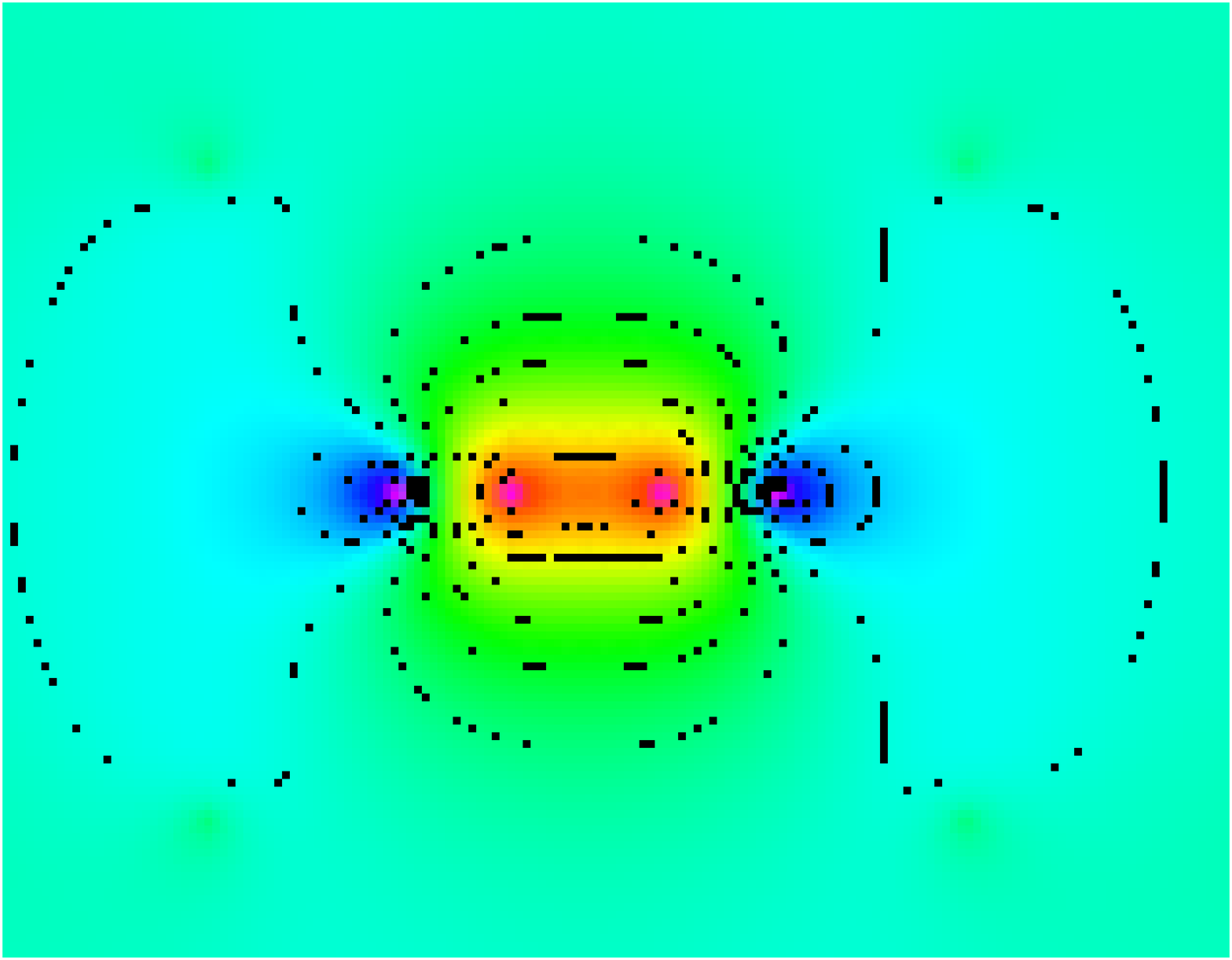}
\caption{\label{hex_c} Contour plot of the FWF1 (left) and MLWF (right)
of an $sp^{2}$-bond of graphene. }
\end{figure}

\begin{figure}
\includegraphics[width=5cm]{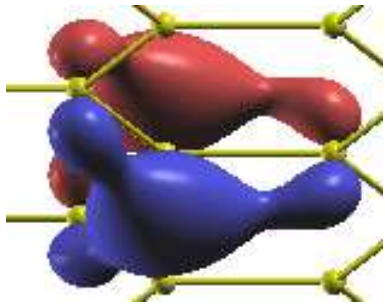}
\caption{\label{hex_c_pi}Isosurface plot of the $\pi$-orbital of
graphene. Isosurface=$\pm$0.1}
\end{figure}

Figure \ref{hex_c} shows the contour plot of one of the three
$sp^{2}$-bonds for the first-guess FWF1 and for the MLWF. 
Figure \ref{hex_c_pi} shows the $\pi$-orbital. Centers  
and spreads are
given in Table
\ref{hex-c-tab}. The initial spread of 17.08 a.u.$^{2}$  
characterizing the first-guess FWF1
is reduced by the minimization procedure to a final total spread of
16.23 a.u.$^{2}$.

The hopping matrix elements $H_{m,m'}(\vn{R}_{1}-\vn{R}_{2})$, Eq.~(\ref{hoppses}),
are listed in Table \ref{hoppings_c}. There is no coupling between the
$\pi$ WFs and the $sp^{2}$ WFs. 

\begin{table}
\caption{\label{hex-c-tab}Centers and spreads of the first-guess
(first row) and maximally localized (second row) WFs (atomic units).}
\begin{ruledtabular}
\begin{tabular}{l|rrrr}
    &$x$   &$y$   &$z$   &$\langle\vn{x}^2\rangle$\\
\hline\\
FWF1 ($sp^{2}$)   &2.038   &1.169  &0.000  & 2.184\\
FWF1 ($sp^{2}$)   &2.038   &$-$1.169 &0.000  & 2.184\\
FWF1 ($sp^{2}$)   &4.064   &0.000  &0.000  & 2.184\\
FWF1 ($\pi$)      &2.714   &0.000  &0.000  &10.526\\
\hline\\
MLWF ($sp^{2}$) &2.035   &1.175  &0.000  & 2.052 \\
MLWF ($sp^{2}$) &2.035   &$-$1.175  &0.000 & 2.052\\
MLWF ($sp^{2}$) &4.070   &0.000  &0.000  & 2.052\\
MLWF ($\pi$)    &2.714   &0.000  &0.000  &10.075\\

\end{tabular}
\end{ruledtabular}
\end{table}

\begin{table}
\caption{\label{hoppings_c}Hopping matrix elements of graphene. Energies are in 
meV. 00, 10, 11 and 20 denote the translations of the  obitals in units of the
primitive translations.}
\begin{ruledtabular}
\begin{tabular}{l|rrrr}
& 00   & 10  & 11  & 20 \\
\hline\\
$        sp^{2} (1) , sp^{2} (1)        $   &-15038 &560.7  &6.6    &51.3  \\
$        sp^{2} (1) , sp^{2} (2)        $   &-2139  &78.0   &-21.5  &7.4   \\
$        sp^{2} (1) , sp^{2} (3)        $   &-2139  &-144.1 &2.5    &-19.9 \\
$        sp^{2} (2) , sp^{2} (1)        $   &-2139  &-529.8 &-21.5  &-21.5 \\
$        sp^{2} (2) , sp^{2} (2)        $   &-15038 &-109.7 &6.6    &-6.7  \\
$        sp^{2} (2) , sp^{2} (3)        $   &-2139  &78.0   &2.5    &7.4   \\
$        sp^{2} (3) , sp^{2} (1)        $   &-2139  &-2139.1&78.0   &-144.1\\
$        sp^{2} (3) , sp^{2} (2)        $   &-2139  &-529.8 &78.0   &-21.5 \\
$        sp^{2} (3) , sp^{2} (3)        $   &-15038 &560.7  &-16.4  &51.3  \\
$        \pi , \pi        $   &-8329   &-728.0  &162.9  & 51.6 \\
\end{tabular}
\end{ruledtabular}
\end{table}

\subsection{Platinum}
\begin{figure}
\includegraphics[width=3.5cm]{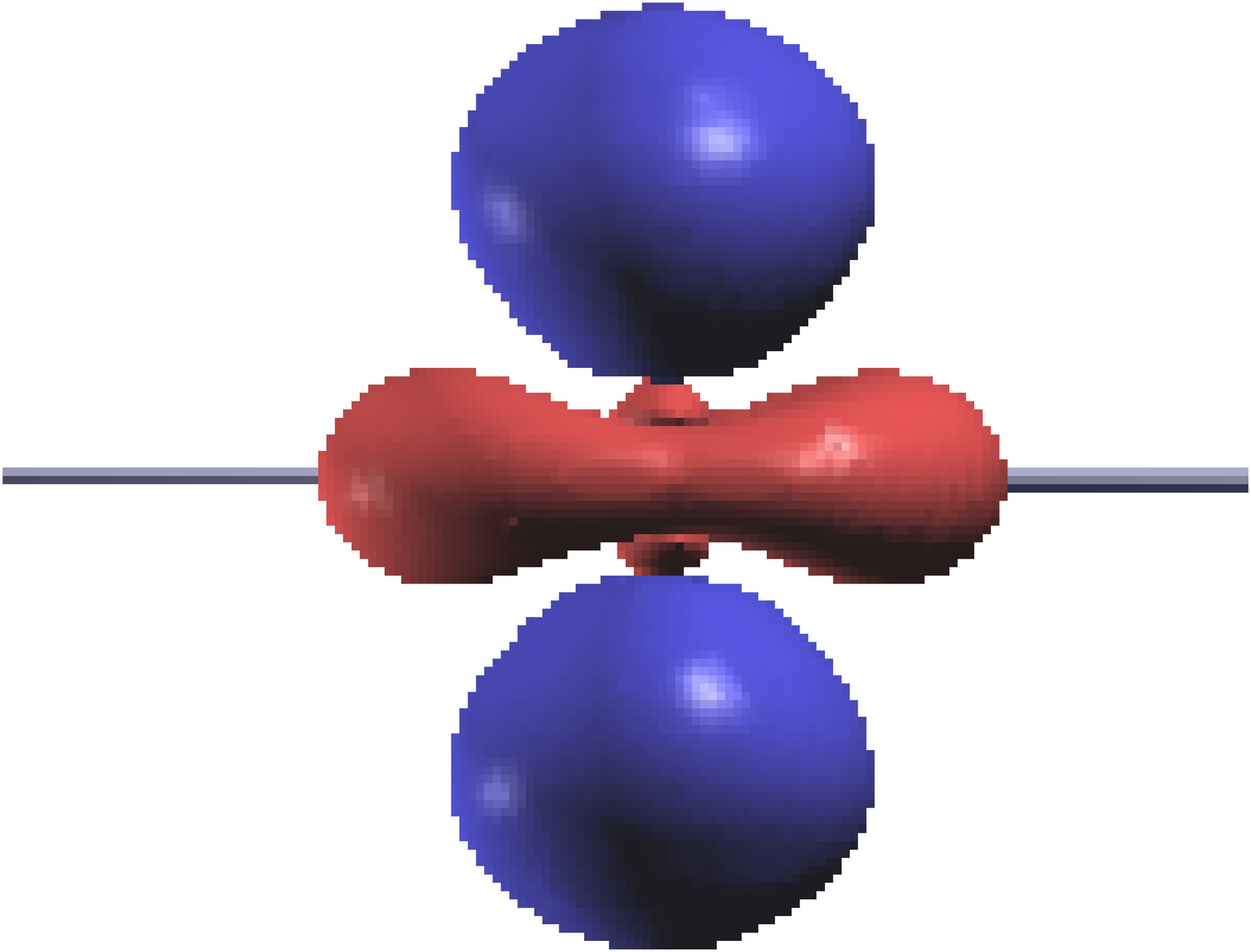}
\includegraphics[width=3.5cm]{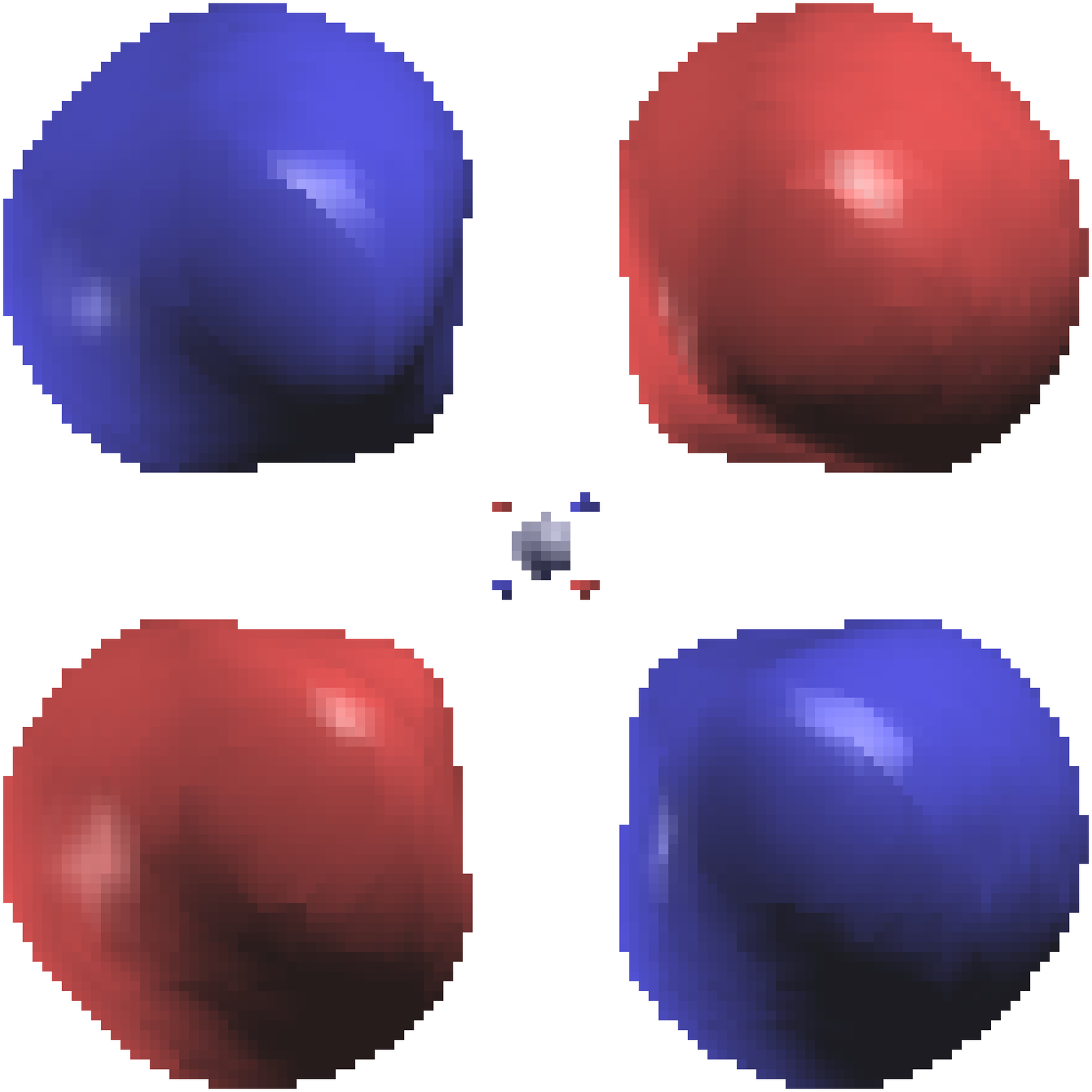}
\includegraphics[width=3.5cm]{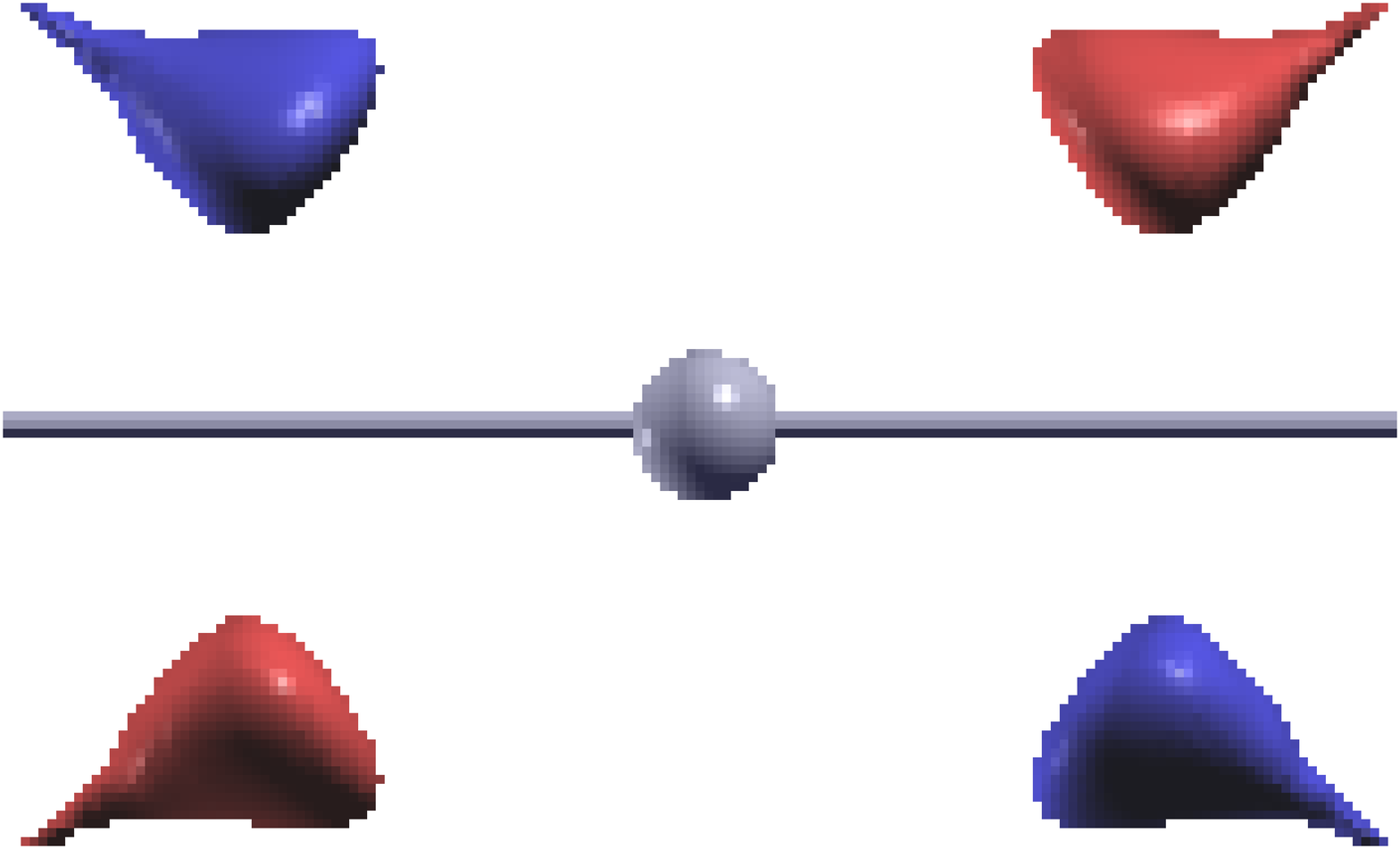}
\includegraphics[width=3.5cm]{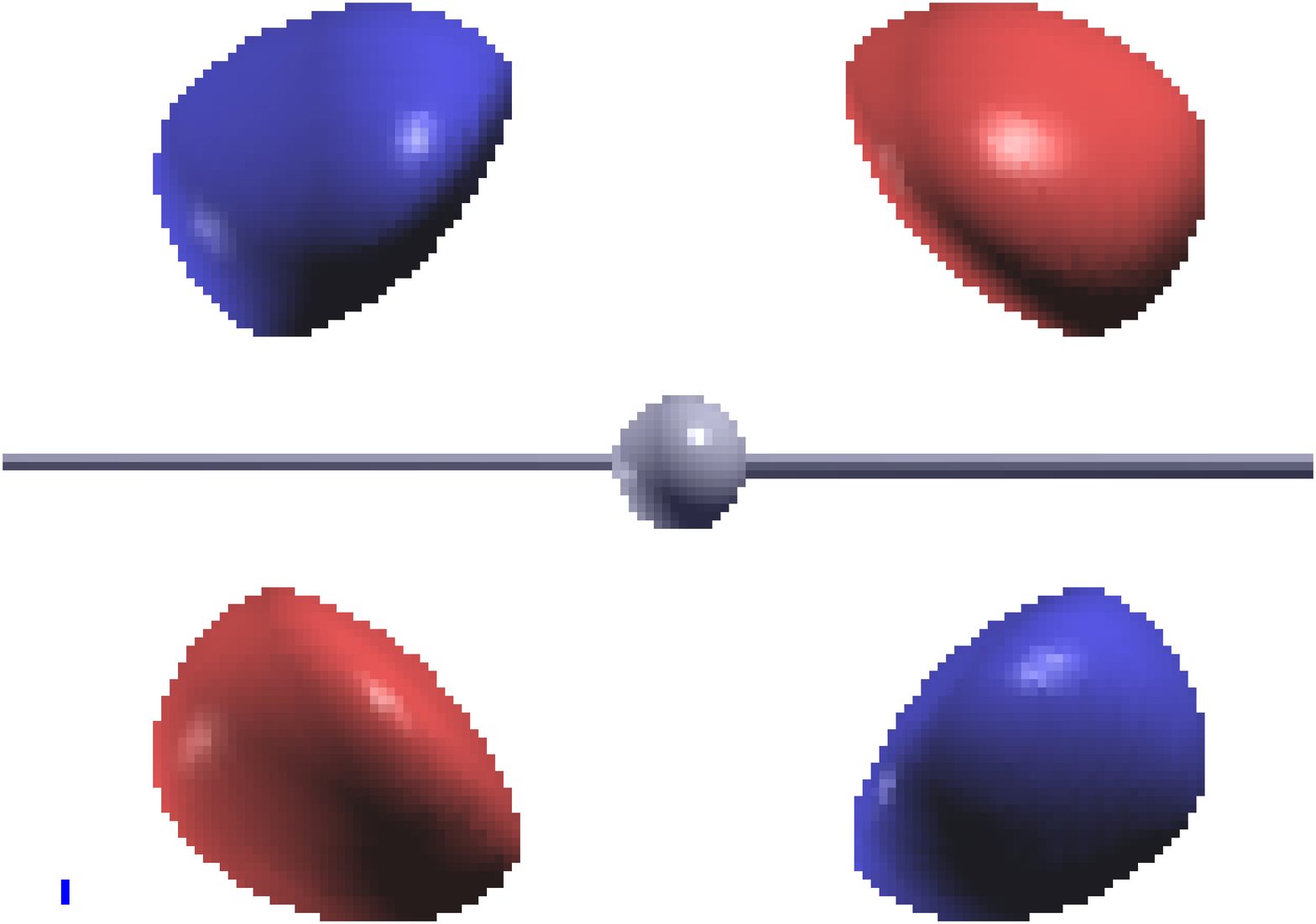}
\caption{\label{pt_y2} $d_{3y^2-r^2}$-like orbital of a one-dimensional Pt-chain. Top row:
Left: Real part of spin-up component ($d_{3y^2-r^2}$, Isosurface=$\pm$0.1), 
Right: imaginary part of spin-up
component ($d_{xy}$, Isosurface=$\pm$0.001). Bottom row: Left: 
Real part of spin-down component
($d_{xz}$, Isosurface=$\pm$0.00073), 
Right: imaginary part of spin-down component 
($d_{yz}$, Isosurface=$\pm$0.0025).}
\end{figure}

\begin{figure}
\includegraphics[width=2.8cm]{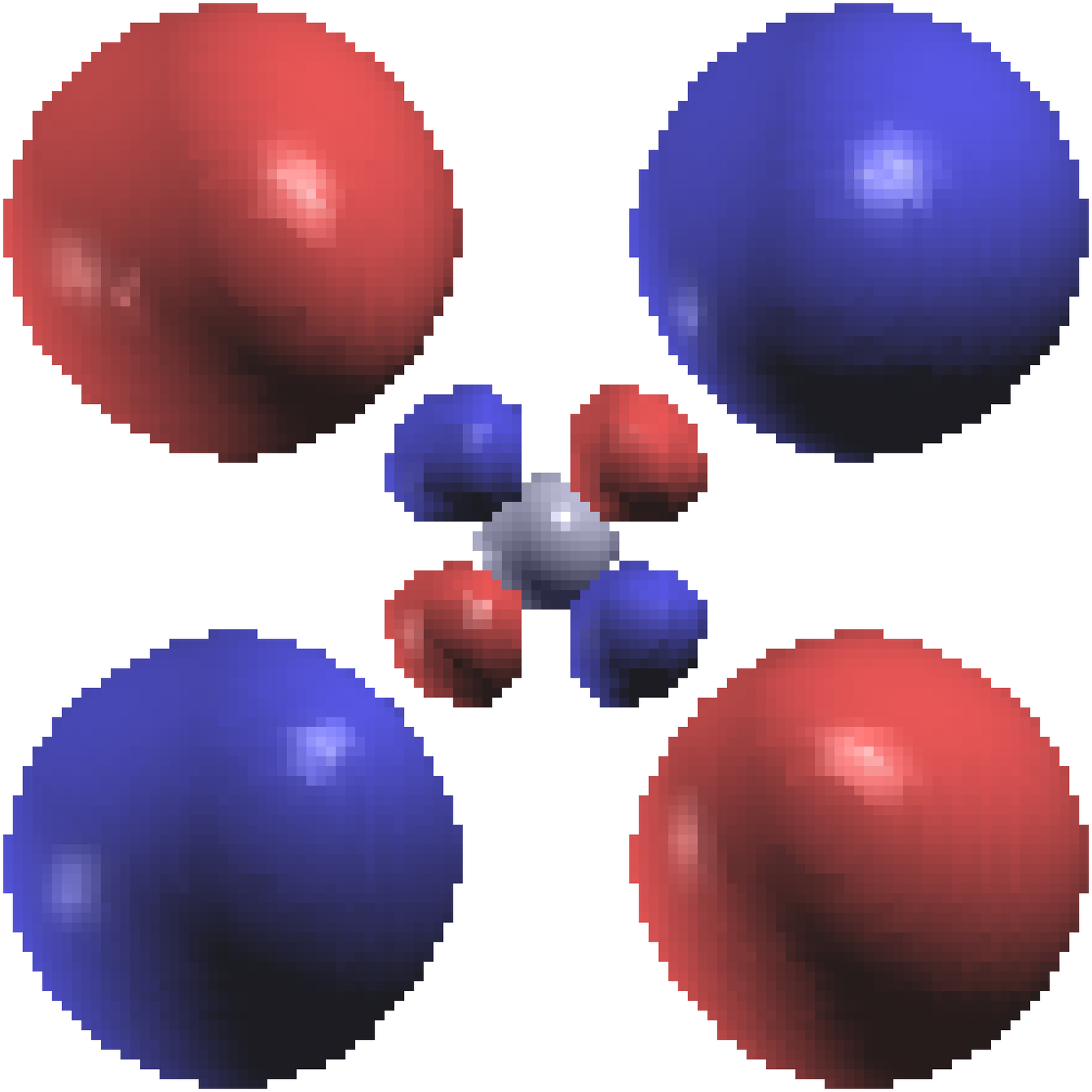}
\includegraphics[width=2.8cm]{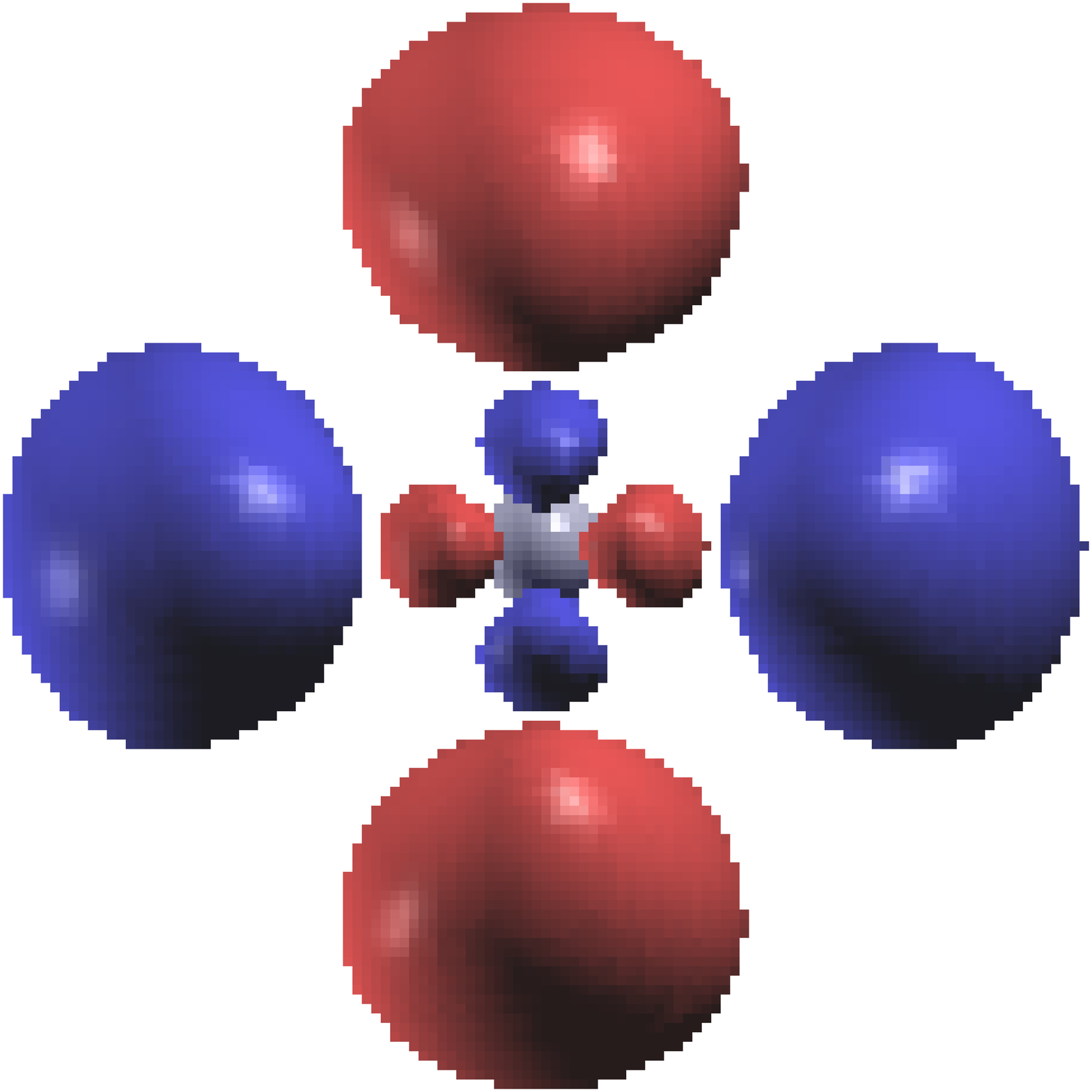}
\includegraphics[width=2.8cm]{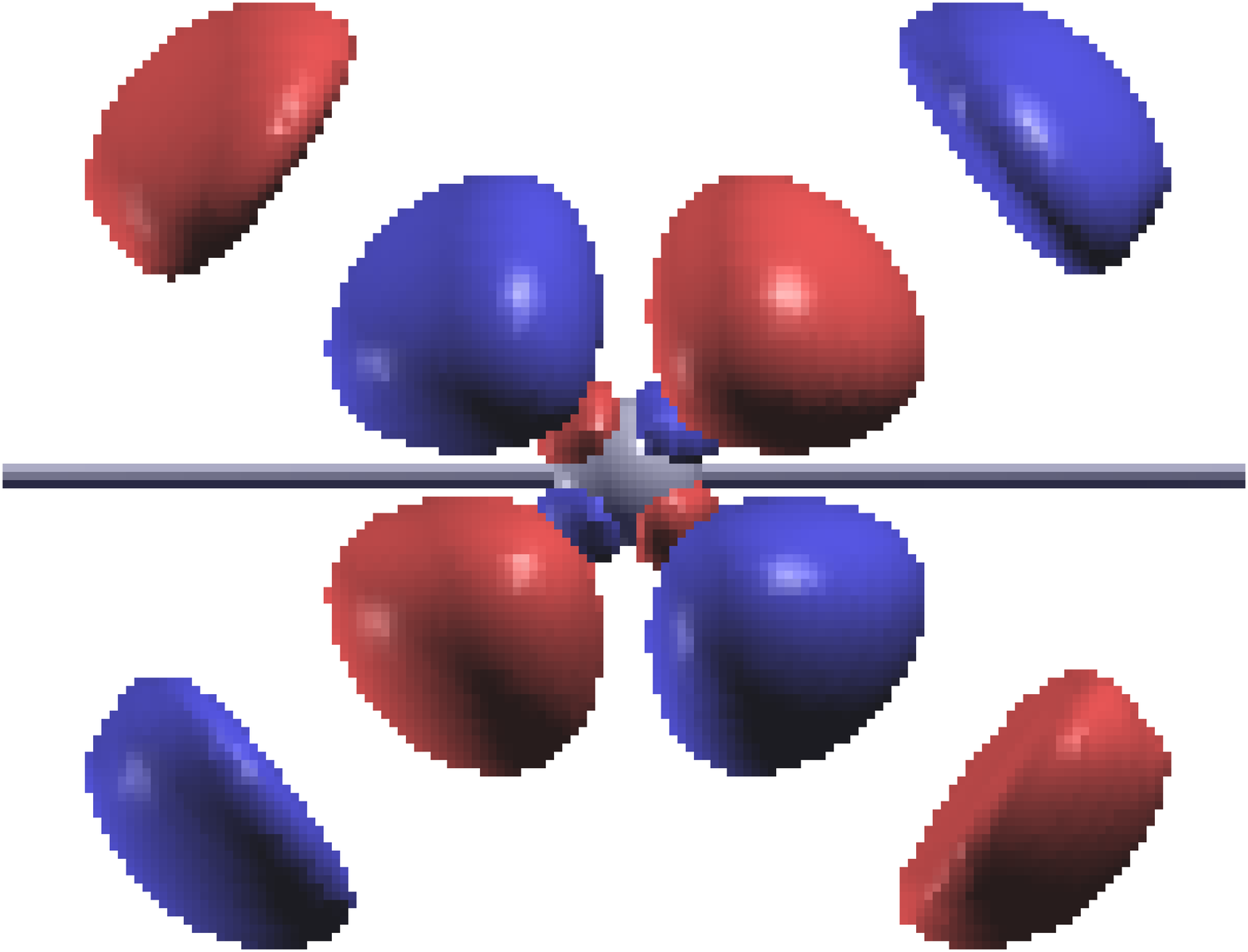}
\caption{\label{pt_xy} $ d_{xy} $-like orbital of a one-dimensional Pt-chain.
From left to right: Real part of spin-up component ($d_{xy}$, Isosurface=$\pm$0.2),
imaginary part of spin-up component ($d_{x^2-y^2}$, Isosurface=$\pm$0.005),
real part of spin-down component ($d_{yz}$, Isosurface=$\pm$0.001).}
\end{figure}

\begin{figure}
\includegraphics[width=3.5cm]{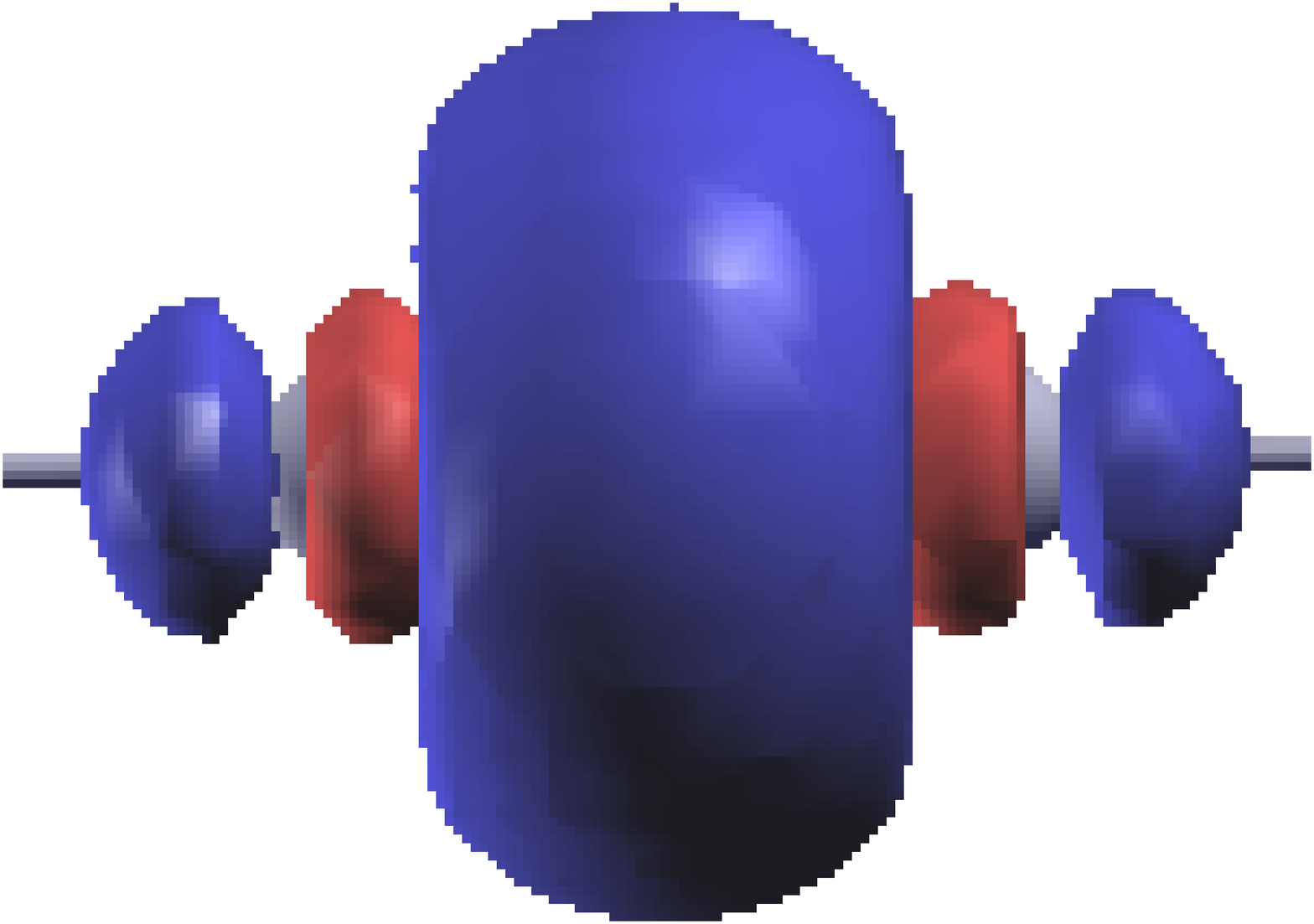}
\includegraphics[width=3.5cm]{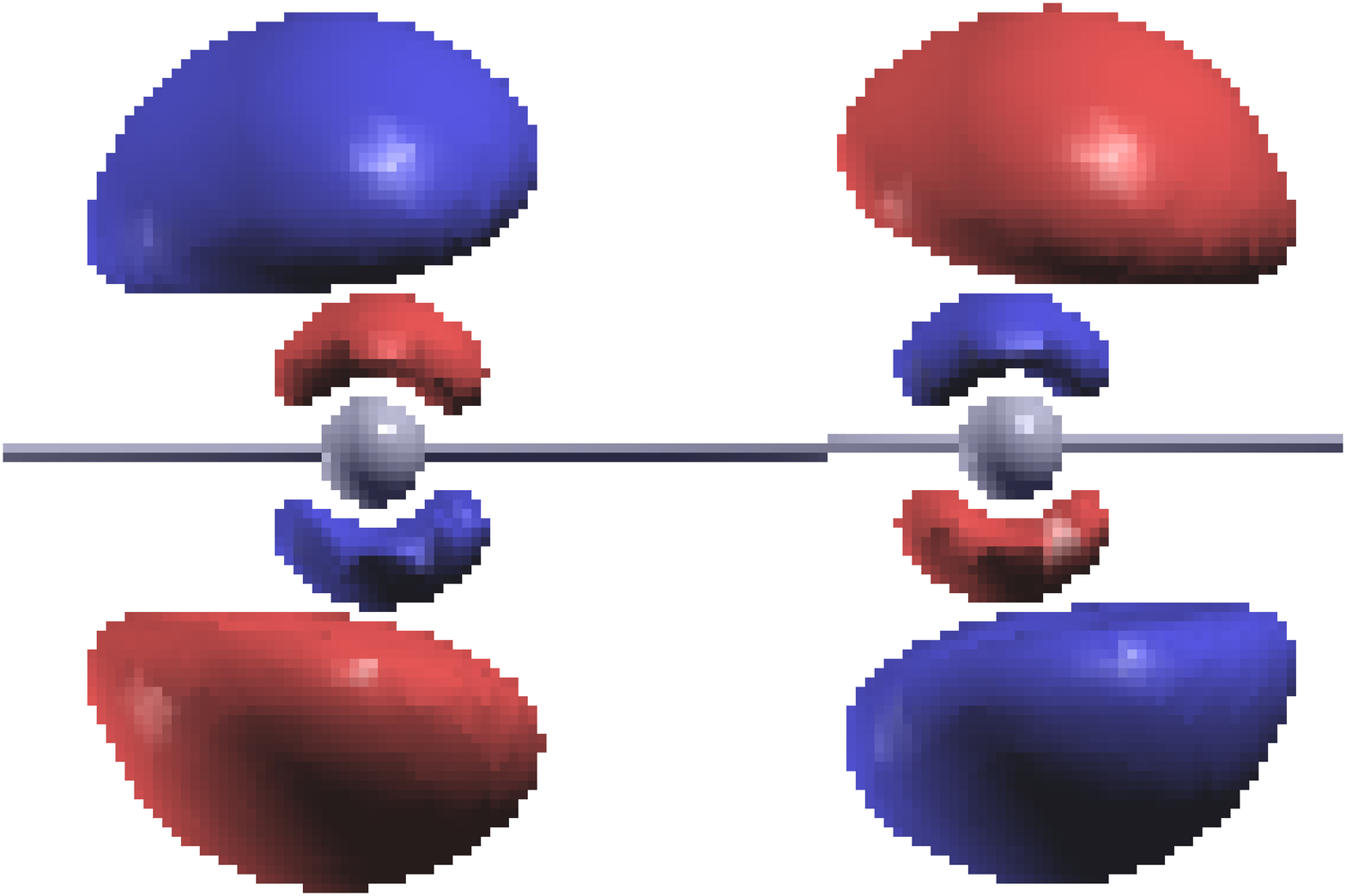}
\caption{\label{pt_sp} $sp$-like orbital of a one-dimensional Pt-chain.
Left: real part of spin-up component ($sp$, Isosurface=$\pm$0.04),
Right: real part of spin-down component ($p_{x}$, Isosurface=$\pm$0.004).}
\end{figure}

We close the results section with the discussion of the MLWFs for a 
Platinum chain. Our calculations were performed with the one-dimensional
version~\cite{trueonedim} of the {\tt FLEUR} program and with spin-orbit 
coupling~\cite{soc-chain1,soc-chain2,soc-chain3,soc-chain4}. 
The extensions necessary to treat the spin-orbit case have been described
in section \ref{spinorbitcoupling}. The muffin-tin 
radii and the plane wave cut-off were chosen to be 2.22 a.u.\ and 3.7 a.u.$^{-1}$, 
respectively. The RPBE~\cite{rpbe-excorr} exchange-correlation potential was used. The
relaxed Pt-Pt distance is given by 4.48 a.u..
We calculated 12 MLWFs corresponding to the $s$- and $d$-states of Platinum using 
8 $k$-points. The localized
trial orbitals were chosen to be eigenstates of the $z$-projection of the spin 
operator. Both the direction of the chain and the spin-quantization axis are
given by the $z$-direction. We chose the angular parts of the trial-orbitals
for the $d$-bands to be $d_{3x^2-r^2}$, $d_{3y^2-r^2}$, 
(i.e. $d_{3z^{2}-r^2}$ rotated to 
be coaxial with the $x$- and $y$-directions, respectively),
$d_{xy}$, $d_{xz}$ and $d_{yz}$. The 
localized trial orbital corresponding to the sp-like 
WF was constructed as
a linear combination of two localized s-orbitals on neighboring atoms. The MLWFs are
spinor-valued and complex. 6 out of the 12 MLWFs are characterized by a dominance of the
spin-up component while the spin-down component dominates the other 6 MLWFs. The two groups
of spin-up and spin-down dominated WFs are symmetric by interchange of spins. Hence we will
consider only the 6 spin-up dominated WFs in the following, unless explicitly stated. The 
angular dependencies of the
real parts of the dominating spin-up components are
approximately given by $d_{xz}$ and $d_{yz}$, $d_{3x^{2}-r^2}$ 
and $d_{3y^{2}-r^2}$, $d_{xy}$, and $sp$.
The MLWFs $d_{xz}$, $d_{yz}$ and $d_{3x^2-r^2}$, $d_{3y^2-r^2}$ are symmetry equivalent, respectively. 
The $sp$-like WF
is positioned bond-centred between two neighboring Pt-atoms. The angular functions that
approximately describe the imaginary part of the spin-up component as well as the real and
imaginary parts of the spin-down components agree very well qualitatively with the results
of our simple
model of section \ref{spinorbitcoupling} given in Table \ref{soc_angular_deps}. We found
qualitative deviations only for the 
$d_{3y^2-r^2}$-orbital (and the symmetry-equivalent $d_{3x^2-r^2}$-orbital) shown in 
Figure \ref{pt_y2}: 
While Table \ref{soc_angular_deps} 
predicts the real part
of the spin-down component belonging to the $d_{3y^2-r^2}$-orbital to vanish, it turns out to be
non-vanishing and $d_{xz}$-like. This may be attributed to the fact that the 
actual $d_{3y^2-r^2}$-like
orbital is not rotationally invariant around the $y$-axis, but rather squeezed in $x$-direction. 
The $d_{xy}$-like WF is shown in Figure \ref{pt_xy}. As there is no spin-orbit coupling for 
$s$-states the spin-down component of the $sp$-like WF, which is shown in 
Figure \ref{pt_sp}, is $p$-like.

Table \ref{spreads_pt} lists the spreads. The maximal localization procedure reduces the initial
total spread of 195.72 a.u.$^{2}$ to a final total spread of 37.56 a.u.$^{2}$.
 
In Table \ref{hoppings_pt} we list the spin-resolved nearest neighbor hopping matrix elements 
for the spin-up dominated MLWFs between identical orbitals calculated according to 
Eq. (\ref{hoppsesspindecomp}). As the $(\downarrow,\downarrow)$ components scale quadratically
with the admixture of spin-down to the spin-up dominated WFs, they are small.
Likewise the
$(\uparrow,\downarrow)$ components are found to be small: The angular distributions 
of the spin-down components of the WFs differ from those of the spin-up components,  the admixture
of spin-down is small, and the spin-orbit coupling, which couples the two spin-channels,
is important only close to the nuclear cores and hence the coupling between 
functions well-localized
on different atoms is small. For the on-site hopping matrix elements, however, 
the $(\uparrow,\downarrow)$- or
$(\downarrow,\uparrow)$-components can dominate, because the two WFs are centered 
on the same atoms
in this case, and their overlap close to the nuclear cores can be large. In Table 
\ref{hoppings_pt_onsite} we list a selection of spin-resolved on-site hopping matrix elements 
that are 
dominated by hopping from spin-up into spin-down, which is mediated by spin-orbit coupling.
$d^{\uparrow}_{xz}$ is a spin-up dominated $d_{xz}$-like WF. 
According to Table \ref{soc_angular_deps}
the spin-orbit interaction provides a coupling to \mbox{$d_{x^2-y^2}|\downarrow\rangle$}, 
which overlaps
with $d^{\downarrow}_{3x^2-r^2}$. Analogously, there is a transition from $d^{\uparrow}_{3y^2-r^2}$ 
to $d_{yz}|\downarrow\rangle$, which overlaps with $d^{\downarrow}_{yz}$. The other two 
examples in Table
\ref{hoppings_pt_onsite} are easily interpreted analogously on the basis of 
Table \ref{soc_angular_deps}.
The $(\downarrow,\uparrow)$-contributions in Table \ref{hoppings_pt_onsite} are 
negligibly small, because
the $|\downarrow\rangle$- and $|\uparrow\rangle$-components of the spin-up and 
spin-down dominated WFs
are small, respectively. Table \ref{hoppings_pt_nn} is analogous to Table \ref{hoppings_pt},
but now for the nearest neighbor hoppings. The comparison of the two Tables shows that the
$(\uparrow,\downarrow)$-contributions decay fastest, which is consistent with the facts that
the spin-orbit coupling is strongest close to the nucleii, and that the WFs are well localized. 
\begin{table}
\caption{\label{spreads_pt}Platinum chain: Spreads of the MLWFs (atomic units).}
\begin{ruledtabular}
\begin{tabular}{l|rrrr}
&$d_{xz}$  & $d_{3x^2-r^2}$ & $d_{xy}$  & $sp$ \\
\hline\\
$\langle \vn{x}^{2} \rangle $   &3.336  & 2.416  &2.326  & 4.952  \\
\end{tabular}
\end{ruledtabular}
\end{table}

\begin{table}
\caption{\label{hoppings_pt}Platinum chain: Spin-resolved nearest neighbor hopping 
matrix elements for the spin-up dominated MLWFs between identical orbitals (meV).}
\begin{ruledtabular}
\begin{tabular}{l|rrrr}
                         &$d_{xz}$, $d_{xz}$  & $d_{3x^2-r^2}$, $d_{3x^2-r^2}$ & $d_{xy}$, $d_{xy}$  & $s$, $s$ \\
\hline\\
$\uparrow,\uparrow $     &1170.9  & -548.8  &-269.7  & -2481.7  \\
$\uparrow,\downarrow$    &-0.1    & 0.4     &-0.1    &     29.3  \\
$\downarrow, \downarrow$ &1.0     &-0.6     &-0.7    &    -21.3\\
\end{tabular}
\end{ruledtabular}
\end{table}

\begin{table}
\caption{\label{hoppings_pt_onsite}Platinum chain: Spin-resolved on-site hopping 
matrix elements between spin-up and spin-down dominated MLWFs (meV).}
\begin{ruledtabular}
\begin{tabular}{l|rrrr}
&$d^{\uparrow}_{xz}$, $d^{\downarrow}_{3x^2-r^2}$  
&$d^{\uparrow}_{3y^2-r^2}$,$d^{\downarrow}_{yz}$ 
&$d^{\uparrow}_{xz}$, $d^{\downarrow}_{xy}$  
&$d^{\uparrow}_{xy}$,$d^{\downarrow}_{xz}$ \\
\hline\\
$\uparrow,\uparrow $     &-142  &134  &10  & -6  \\
$\uparrow,\downarrow$    &460   &460  &268 & 268  \\
$\downarrow,\uparrow$    &0     &0    &0   & 0   \\
$\downarrow, \downarrow$ &134   &-142 &-6  & 10   \\
\end{tabular}
\end{ruledtabular}
\end{table}

\begin{table}
\caption{\label{hoppings_pt_nn}Platinum chain: Spin-resolved nearest neighbor hopping 
matrix elements between spin-up and spin-down dominated MLWFs (meV).}
\begin{ruledtabular}
\begin{tabular}{l|rrrr}
&$d^{\uparrow}_{xz}$, $d^{\downarrow}_{3x^2-r^2}$  
&$d^{\uparrow}_{3y^2-r^2}$,$d^{\downarrow}_{yz}$ 
&$d^{\uparrow}_{xz}$, $d^{\downarrow}_{xy}$  
&$d^{\uparrow}_{xy}$,$d^{\downarrow}_{xz}$ \\
\hline\\
$\uparrow,\uparrow $     &33  &0.8  &5.6  & -9.0  \\
$\uparrow,\downarrow$    &9.8   &9.8  &7.5 & 7.5  \\
$\downarrow,\uparrow$    &0     &0    &0   & 0   \\
$\downarrow, \downarrow$ &0.8   &33 &-9.0  & 5.6   \\
\end{tabular}
\end{ruledtabular}
\end{table}

\section{Conclusions}
We have described the implementation of Wannier functions
within the FLAPW program {\tt FLEUR} for bulk, film and wire
geometry. Two kinds of WFs with optimized
localization properties -- the first-guess and the maximally localized
Wannier functions -- have been described and calculated for four 
concrete systems, SrVO$_{3}$, BaTiO$_{3}$, graphene and platinum. Our results
are in very good agreement to previous ones, where available, including
the ferroelectric polarization of BaTiO$_{3}$.
We found the
first-guess WFs and the MLWFs to be similar for the first three systems, and
rather different for Pt. While in
cases where the first-guess WFs and the MLWFs do not
differ substantially there is the option to use the first-guess WFs
in practice for certain applications, which is computationally less 
demanding, the extended scheme needed for the construction of the 
MLWFs still proves valuable if quantities such as the electric 
polarization are supposed to be extracted.  
\acknowledgments
We thank Eva Pavarini and Gustav Bihlmayer for
fruitful discussions.
Financial support of the Stifterverband f\"ur die Deutsche
Wissenschaft and the Interdisciplinary Nanoscience
Center Hamburg are gratefully acknowledged.

\appendix

\section{Vacuum contributions to the $M_{mn}^{(\vn{k},\vn{b})}$ matrix in case of film
calculations}
\label{vacappendfilm}
In case of the film implementation of the FLAPW
method, an additional semi-infinite vacuum region is present, which  
results in an
additional contribution to the wave function overlaps
$M_{mn}^{(\vn{k},\vn{b})}|_{\text{VAC}}$. In this appendix we give  
explicit
expressions for the vacuum contributions to the $M_{mn}^{(\vn{k},\vn 
{b})}$
matrix elements.

In the film geometry, the interstitial region stretches in z-direction from 
$-D/2$ to $D/2$, 
which is chosen to be the direction orthogonal to the film.
Thus, one of the two vacua extends from $-\infty$ to $-D/2$ while
the second vacuum extends from $D/2$ to $+\infty$. The two
vacua are treated analogously and we will restrict the discussion
to the vacuum between $D/2$ and $+\infty$.
According to the topology of the vacuum region, the Bloch wave
functions in the vacuum are represented in the following way:
\begin{equation}
\label{A1}
\displaystyle  \psi_{\vn{k_{\Vert}}m}(\vn{x})|_{\text{VAC}} =
  \sum_{\vn{G_{\Vert}}}\Psi^{m}_{\vn{G_{\Vert}}}(\vn{k_{\Vert}},z)
   e^{i(\vn{G_{\Vert}}+\vn{k_{\Vert}})\cdot\vn{x}_{\Vert}},
\end{equation}
with
\begin{equation}
\label{A2}
\Psi^{m}_{\vn{G_{\Vert}}}(\vn{k}_{\Vert},z)=A^{m}_{\vn{G_{\Vert}}}(\vn{k}_{\Vert})
   u_{\vn{G_{\Vert}}}^{\vn{k}_{\Vert}}(z)+
   B^{m}_{\vn{G_{\Vert}}}(\vn{k}_{\Vert})
   \dot{u}_{\vn{G_{\Vert}}}^{\vn{k}_{\Vert}}(z),
\end{equation}
where
$\vn{G}=(\vn{G_{\Vert}},G_z)$ and
$\vn{x}=(\vn{x}_{\Vert},z)$ have been used, with $\vn{G_{\Vert}}$
and $\vn{x_{\Vert}}$ the in-plane components.
The $k$-point $\vn{k}_{\Vert}$ belongs to the
two-dimensional BZ. $u_{\vn{G_{\Vert}}}^{\vn{k}}(z)$ and
$\dot{u}_{\vn{G_{\Vert}}}^{\vn{k}}(z)$ are the solution of
the one-dimensional Schr\"odinger equation in the vacuum and its energy
derivative, respectively.
Substituting Eq.~\ref{A1} into Eq.~\ref{mmnwhatis} yields:
\begin{equation}
\label{mmnvac}
\begin{aligned}
&M_{mn}^{(\vn{k}_{\Vert},\vn{b})}\\
=\sum_{\vn{G_{\Vert}},\vn{G^{\prime}_{\Vert}}} 
\displaystyle
\int_{\text{VAC}} &e^{i\vht{\mathcal{G}}\cdot\vn{x}}
(\Psi^{m}_{\vn{G_{\Vert}}}(\vn{k}_{\Vert},z))^{*}\Psi^{n}_{\vn{G^{\prime}_ 
{\Vert}}}([\vn{k}_{\Vert}+\vn{b}],z)\,d^3x
\end{aligned}
\end{equation}	
with $\vht{\mathcal{G}}=\vn{G^{\prime}_{\Vert}}-\vn{G_{\Vert}}-\vn{G} 
(\vn{k}_{\Vert}+\vn{b})$. While vectors 
$\vn{k}_{\Vert}$ and $[\vn{k}_{\Vert}+\vn{b}]$ always lie in the 
two-dimensional Brillouin zone, 
the $ \vn{b}$ and $\vn{G}(\vn{k}_{\Vert}+\vn{b})$
vectors have a $z$-component in general, which leads to the following  
expression for the
$M_{mn}^{(\vn{k}_{\Vert},\vn{b})}$ matrix elements:
\begin{equation}
\label{mmnvac2}
\begin{aligned}
&M_{mn}^{(\vn{k}_{\Vert},\vn{b})}=\sum_{\vn{G_{\Vert}},\vn{G^ 
{\prime}_{\Vert}}}
S_{\Vert}\delta_{\vht{\mathcal{G}}_{\Vert}} \\
\times\int^{\infty}_{D/2}& e^{-i\,G_{z}(\vn{k}_{\Vert}+\vn{b})\, z}
(\Psi^{m}_{\vn{G_{\Vert}}}(\vn{k}_{\Vert},z))^{*}\Psi^{n}_{\vn{G^{\prime}_ 
{\Vert}}}([\vn{k}_{\Vert}+\vn{b}],z)\,dz,
\end{aligned}
\end{equation}
with $S_{\Vert}$ being the in-plane unit-cell area, and the last  
integral is a linear
combination of one-dimensional integrals of the form
\begin{equation}
\begin{aligned}
&\int^{\infty}_{D/2} e^{-i\,G_z(\vn{k}_{\Vert}+\vn{b})\, z}\,u_{\vn{G_{\Vert}}} 
^{\vn{k}_{\Vert}}(z)\,u_{\vn{G^{\prime}_{\Vert}}}^{[\vn{k}_{\Vert}+\vn{b}]}(z)\,dz,\\
&\int^{\infty}_{D/2} e^{-i\,G_z(\vn{k}_{\Vert}+\vn{b})\, z}\,u_{\vn{G_{\Vert}}} 
^{\vn{k}_{\Vert}}(z)\,\dot{u}_{\vn{G^{\prime}_{\Vert}}}^{[\vn{k}_{\Vert}+\vn{b}]}(z)\,dz,\\
\end{aligned}
\end{equation}
which are easily computed numerically for every pair of
$(\vn{G_{\Vert}},\vn{G^{\prime}_{\Vert}})$.

\section{Vacuum contributions to the $M_{mn}^{(\vn{k},\vn{b})}$ matrix in case of one
dimensional calculations}
\label{vacappendonedim}

In the case of the one-dimensional setup the vacuum region surrounds
a cylinder with the symmetry axis along the $z$-direction and radius
$R_{\text{vac}}$. The wave function in the vacuum is represented in
the following form (in the 1D case the Bloch vector is $\vn{k}=(0,0,k_z)$):
\begin{equation}
\begin{array}{cc}
\displaystyle\psi_{k_{z} m}(\vn{x}) = \sum_{G_z,p}(A_{p,G_z}^{m,k_{z}}u_p^{G_z}(k_z,r)
+ B_{p,G_z}^{m,k_{z}}\dot{u}_p^{G_z}(k_z,r))\times \\[0.3cm]
\displaystyle\times e^{ip\varphi}e^{i(G_z+k_z)z},
\end{array}
\end{equation}
where $\vn{x}=(z,r,\varphi)$ in cylindrical coordinates,
$G_z$ is the $z$-component
of the reciprocal vector $\vn{G}$, and $p$ is an integer number labeling
a cylindrical
angular harmonic. The exponentially decaying functions $u$ and $\dot{u}$
are the solutions of the radial equation for the vacuum and its
energy derivative, respectively. Taking into account the
expansion of a plane wave in cylindrical coordinates
\begin{equation}
e^{i\vn{G}\vn{x}}=e^{iG_z z}\sum_p \,i^{p}e^{ip(\varphi-\varphi_{\vn{G}})}J_p(G_r r),
\end{equation}
with $\varphi_{\vn{G}}$ and $G_r$ being cylindrical angular and radial coordinates,
respectively, of
the vector $\vn{G}=(G_z,G_r,\varphi_{\vn{G}})$ in reciprocal space,
and $J_p$ standing for the
cylindrical Bessel function of order $p$, the 1D-vacuum contribution to the
$M_{mn}^{(k_{z},\vn{b})}$ matrix reads:
\begin{equation}
\begin{array}{cc}
\displaystyle M_{mn}^{(k_{z},\vn{b})}|_{\text{VAC}}=
\int_{\text{VAC}} e^{-i\vn{b}\cdot\vn{x}}(\psi_{k_{z}m}(\vn{x}))^{*}\psi_{[k_{z}+\vn{b}],n}
(\vn{x})\,d^{3}x \\[0.5cm]
=\displaystyle\sum_{G_z,G^{\prime}_z}\sum_{p,p^{\prime}}\int_{\text{VAC}}
e^{i(G^{'}_z-G_z-G_z(k_{z}+\vn{b}))z}\times\\[0.7cm]
\displaystyle\times \,e^{-i\vn{G}_{\Vert}(k_{z}+\vn{b})\vn{x}_{\Vert}}
\,e^{i(p^{\prime}-p)\varphi}\,\Psi_{p,p^{\prime},G^{\prime}_z}^{m,n,G_z}(k_{z},[k_{z}+\vn{b}],r)\,d^{3}x,
\end{array}
\end{equation}
where in analogy to the case of the film geometry, vectors $\vn{b}$ and
$\vn{G}(k_{z}+\vn{b})$ may have a non-zero component in the plane
normal to the $z$-axis, and the function
$\Psi$ is constructed from the products of the $u$- and
$\dot{u}$-functions with corresponding $A$- and $B$-coefficients at $k$-points
$k_{z}$ and $[k_{z}+\vn{b}]$.
Introducing the vector $\mathcal{G}=G^{'}_z-G_z-G_z(k_{z}+\vn{b})$ the expression
for the $M_{mn}^{(k_{z},\vn{b})}$ can be reduced to
\begin{equation}
\begin{array}{cc}
\displaystyle M_{mn}^{(k_{z},\vn{b})}|_{\text{VAC}}=
\sum_{G_z,G^{'}_z}\sum_{p,p^{\prime}}\mathcal{S}\cdot \delta_{\mathcal{G}}
\cdot i^{p-p^{\prime}}e^{-i(p-p^{\prime})\varphi_{\vn{G}(k_{z}+\vn{b})}}
\times\\[0.5cm]
\displaystyle\times\int_{R_{\text{vac}}}^{\infty}
r J_{p^{\prime}-p}(G_r(k_{z}+\vn{b})r)\Psi_{p,p^{\prime},G^{\prime}_z}^{m,n,G_z}(k_{z},[k_{z}+\vn{b}],r)\,dr,
\end{array}
\end{equation}
with $\mathcal{S}=2\pi T$, and $T$ standing for the lattice constant of the system
under consideration along the $z$-axis.

\section{Local orbital contributions to the $M_{mn}^{(\vn{k},\vn{b})} 
$ matrix}
\label{locorbs}
In order to increase the variational freedom of the FLAPW-basis or to
describe semicore levels adequately, it may
be supplemented by local orbitals.~\cite{Singh}
In this case the expressions for the
BFs in the spheres are modified:
\begin{equation}\label{bflo}
\begin{array}{cc}
   \displaystyle\psi_{\vn{k}m}(\vn{x})|_{\text{MT}^{\mu}} =
            \sum_{L}(A^{\mu}_{L,m}(\vn{k})
   u_{l}^{\mu}(r)+B^{\mu}_{L,m}(\vn{k})
   \dot{u}_{l}^{\mu}(r))
   Y_{L}(\hat{\vn{r}})\\
\displaystyle+\sum_{Lo}C^{\mu}_{Lo,m}(\vn{k})
   u_{lo}^{\mu}(r)Y_{Lo}(\hat{\vn{r}}),
\end{array}
\end{equation}
where $Lo=(lo,mo)$ stands for the corresponding values of the angular
quantum numbers $(l,m)$ assigned to each local orbital. Due to the local
orbitals, additional terms arise
in the expression Eq.~\ref{mmn_sum} for the
$M_{mn}^{(\vn{k},\vn{b})}|_{\text{MT}^{\mu}}$ matrix:
\begin{equation}
\begin{array}{cc}
M_{mn}^{(\vn{k},\vn{b})}|^{Lo}_{\text{MT}^{\mu}}=4\pi e^{-i\vn{b}\cdot 
\vht{\tau}_{\mu}}\times\\[0.2cm]	
\displaystyle\times(\sum_{L,Lo^{\prime}}(A^{\mu}_{L,m}(\vn{k}))^{*}C^ 
{\mu}_{Lo',m}(\vn{[k+b]})\,
                                    t_{11}^{\mu}(\vn{b},L,Lo') + \\
\displaystyle+\sum_{L,Lo^{\prime}}(B^{\mu}_{L,m}(\vn{k}))^{*}C^{\mu}_ 
{Lo',m}(\vn{[k+b]})\,
                                    t_{21}^{\mu}(\vn{b},L,Lo') + \\
\displaystyle+\sum_{Lo,L'}(C^{\mu}_{Lo,m}(\vn{k}))^{*}A^{\mu}_{L^ 
{\prime},m}(\vn{[k+b]})\,
                                    t_{11}^{\mu}(\vn{b},Lo,L') + \\
\displaystyle+\sum_{Lo,L'}(C^{\mu}_{Lo,m}(\vn{k}))^{*}B^{\mu}_{L^ 
{\prime},m}(\vn{[k+b]})\,
                                    t_{12}^{\mu}(\vn{b},Lo,L') + \\
\displaystyle+\sum_{Lo,Lo'}(C^{\mu}_{{Lo},m}(\vn{k}))^{*}C^{\mu}_ 
{Lo',m}(\vn{[k+b]})\,
                                    t_{11}^{\mu}(\vn{b},Lo,Lo')),
\end{array}
\end{equation}
where the corresponding radial function for the local orbital is taken
in the $t_{ij}^{\mu}$-integrals, whenever a radial function
$u$ has an index $lo$.

\bibliography{method_wannier}

\begin{thebibliography}{41}
\bibitem[{rea()}]{realitynote}
\bibinfo{note}{In the case of spin-orbit coupling, for example, the requirement
  of real-valuedness of the Wannier function cannot be fulfilled in general.}
\end{thebibliography}

\end{document}